\documentstyle[10pt,epsf,epsfig,dp_delphititle,lineno]{dp_delphi}
\makeindex
\def\DpPaperGroup{EP}
\def\DpPaperRef{2003-071}
\def\DpDate{4 November 2003}
\def\DpAuthors{DELPHI Collaboration}
\def\DpSubmit{(Accepted by Eur. Phys. J. C)}
\def\DpTitle{Measurement of the W-pair Production Cross-section \\
  and W Branching Ratios 
  in $e^+e^-$ Collisions at $\sqrt{s}$ = 161-209 GeV}
\def\DpComment{ }
\def\DpEMail{ }

\newcommand{\rww} {{\cal{R}}_{\mathrm{WW}}}
\newcommand{\W} {\mathrm{W}}
\newcommand{\Z} {\mathrm{Z}}
\newcommand{\RacoonWW} {\mbox{R{\sc acoon}WW}}
\newcommand{\Gentle} {\mbox{G{\sc entle}}}
\newcommand{\ARIADNE} {\mbox{ARIADNE}}
\newcommand{\HERWIG} {\mbox{HERWIG}}
\newcommand{\WPHACT} {\mbox{WPHACT}}
\newcommand{\PYTHIA} {\mbox{PYTHIA}}
\newcommand{\BHWIDE} {\mbox{B{\sc HWIDE}}}
\newcommand{\YFSWW} {\mbox{YFSWW}}
\newcommand{\KoralW} {\mbox{K{\sc oral}W}}
\newcommand{\KoralZ} {\mbox{K{\sc oral}Z}}

\newcommand{\qq} {\mbox{${q\bar{q}}$}}
\newcommand{\qqga} {\mbox{${q\bar{q}(\gamma)}$}}

\newcommand{\qqenu} {\mbox{$\qq e\nu$}}
\newcommand{\qqmunu} {\mbox{$\qq\mu\nu$}}
\newcommand{\qqtaunu} {\mbox{$\qq\tau\nu$}}
\newcommand{\qqlnu} {\mbox{$\qq l \nu$}}
\newcommand{\lnulnu} {\mbox{$l \nu l \nu$}}

\newcommand{\jjenu} {\mbox{$jje\nu$}}
\newcommand{\jjmunu} {\mbox{$jj\mu\nu$}}
\newcommand{\jjtaunu} {\mbox{$jj\tau\nu$}}
\newcommand{\jjjj} {\mbox{$jjjj$}}
\newcommand{\ra}{\rightarrow}
\begin{document}
\makeatletter
\newcount\@tempcntc
\def\@citex[#1]#2{\if@filesw\immediate\write\@auxout{\string\citation{#2}}\fi
  \@tempcnta\z@\@tempcntb\m@ne\def\@citea{}\@cite{\@for\@citeb:=#2\do
    {\@ifundefined
       {b@\@citeb}{\@citeo\@tempcntb\m@ne\@citea\def\@citea{,}{\bf ?}\@warning
       {Citation `\@citeb' on page \thepage \space undefined}}%
    {\setbox\z@\hbox{\global\@tempcntc0\csname b@\@citeb\endcsname\relax}%
     \ifnum\@tempcntc=\z@ \@citeo\@tempcntb\m@ne
       \@citea\def\@citea{,}\hbox{\csname b@\@citeb\endcsname}%
     \else
      \advance\@tempcntb\@ne
      \ifnum\@tempcntb=\@tempcntc
      \else\advance\@tempcntb\m@ne\@citeo
      \@tempcnta\@tempcntc\@tempcntb\@tempcntc\fi\fi}}\@citeo}{#1}}
\def\@citeo{\ifnum\@tempcnta>\@tempcntb\else\@citea\def\@citea{,}%
  \ifnum\@tempcnta=\@tempcntb\the\@tempcnta\else
   {\advance\@tempcnta\@ne\ifnum\@tempcnta=\@tempcntb \else \def\@citea{--}\fi
    \advance\@tempcnta\m@ne\the\@tempcnta\@citea\the\@tempcntb}\fi\fi}
 
\makeatother
\begin{titlepage}
\pagenumbering{roman}
\CERNpreprint{\DpPaperGroup}{\DpPaperRef} 
\date{{\small\DpDate}} 
\title{\DpTitle} 
\address{\DpAuthors} 
\begin{shortabs} 
\noindent
%
\noindent

These final results on $e^+e^-\rightarrow W^+W^-$ production
cross-section measurements at LEP2 use data 
collected by the DELPHI detector at centre-of-mass energies up
to 209 GeV. Measurements of total cross-sections,
$W$ angular differential distributions
and decay branching fractions, and the value of the CKM element
$|V_{cs}|$ are 
compared to the expectations of the Standard Model. 

These results supersede all values previously published
by DELPHI.




\end{shortabs}
\vfill
\begin{center}
\DpSubmit \ \\ 
\DpComment \ \\
\DpEMail \ \\
\end{center}
\vfill
\clearpage
\headsep 10.0pt
\addtolength{\textheight}{10mm}
\addtolength{\footskip}{-5mm}
\begingroup
%
\newcommand{\DpName}[2]{\hbox{#1$^{\ref{#2}}$},\hfill}
\newcommand{\DpNameTwo}[3]{\hbox{#1$^{\ref{#2},\ref{#3}}$},\hfill}
\newcommand{\DpNameThree}[4]{\hbox{#1$^{\ref{#2},\ref{#3},\ref{#4}}$},\hfill}
\newskip\Bigfill \Bigfill = 0pt plus 1000fill
\newcommand{\DpNameLast}[2]{\hbox{#1$^{\ref{#2}}$}\hspace{\Bigfill}}
%
\footnotesize
\noindent
\DpName{J.Abdallah}{LPNHE}
\DpName{P.Abreu}{LIP}
\DpName{W.Adam}{VIENNA}
\DpName{P.Adzic}{DEMOKRITOS}
\DpName{T.Albrecht}{KARLSRUHE}
\DpName{T.Alderweireld}{AIM}
\DpName{R.Alemany-Fernandez}{CERN}
\DpName{T.Allmendinger}{KARLSRUHE}
\DpName{P.P.Allport}{LIVERPOOL}
\DpName{U.Amaldi}{MILANO2}
% ...
\normalsize
\endgroup
\titlefoot{Department of Physics and Astronomy, Iowa State
     University, Ames IA 50011-3160, USA
    \label{AMES}}
\titlefoot{Physics Department, Universiteit Antwerpen,
     Universiteitsplein 1, B-2610 Antwerpen, Belgium \\
     \indent~~and IIHE, ULB-VUB,
     Pleinlaan 2, B-1050 Brussels, Belgium \\
     \indent~~and Facult\'e des Sciences,
     Univ. de l'Etat Mons, Av. Maistriau 19, B-7000 Mons, Belgium
    \label{AIM}}
\titlefoot{Physics Laboratory, University of Athens, Solonos Str.
     104, GR-10680 Athens, Greece
    \label{ATHENS}}
\titlefoot{Department of Physics, University of Bergen,
     All\'egaten 55, NO-5007 Bergen, Norway
    \label{BERGEN}}
\titlefoot{Dipartimento di Fisica, Universit\`a di Bologna and INFN,
     Via Irnerio 46, IT-40126 Bologna, Italy
    \label{BOLOGNA}}
\titlefoot{Centro Brasileiro de Pesquisas F\'{\i}sicas, rua Xavier Sigaud 150,
     BR-22290 Rio de Janeiro, Brazil \\
     \indent~~and Depto. de F\'{\i}sica, Pont. Univ. Cat\'olica,
     C.P. 38071 BR-22453 Rio de Janeiro, Brazil \\
     \indent~~and Inst. de F\'{\i}sica, Univ. Estadual do Rio de Janeiro,
     rua S\~{a}o Francisco Xavier 524, Rio de Janeiro, Brazil
    \label{BRASIL}}
\titlefoot{Coll\`ege de France, Lab. de Physique Corpusculaire, IN2P3-CNRS,
     FR-75231 Paris Cedex 05, France
    \label{CDF}}
\titlefoot{CERN, CH-1211 Geneva 23, Switzerland
    \label{CERN}}
\titlefoot{Institut de Recherches Subatomiques, IN2P3 - CNRS/ULP - BP20,
     FR-67037 Strasbourg Cedex, France
    \label{CRN}}
\titlefoot{Now at DESY-Zeuthen, Platanenallee 6, D-15735 Zeuthen, Germany
    \label{DESY}}
\titlefoot{Institute of Nuclear Physics, N.C.S.R. Demokritos,
     P.O. Box 60228, GR-15310 Athens, Greece
    \label{DEMOKRITOS}}
\titlefoot{FZU, Inst. of Phys. of the C.A.S. High Energy Physics Division,
     Na Slovance 2, CZ-180 40, Praha 8, Czech Republic
    \label{FZU}}
\titlefoot{Dipartimento di Fisica, Universit\`a di Genova and INFN,
     Via Dodecaneso 33, IT-16146 Genova, Italy
    \label{GENOVA}}
\titlefoot{Institut des Sciences Nucl\'eaires, IN2P3-CNRS, Universit\'e
     de Grenoble 1, FR-38026 Grenoble Cedex, France
    \label{GRENOBLE}}
\titlefoot{Helsinki Institute of Physics, P.O. Box 64,
     FIN-00014 University of Helsinki, Finland
    \label{HELSINKI}}
\titlefoot{Joint Institute for Nuclear Research, Dubna, Head Post
     Office, P.O. Box 79, RU-101 000 Moscow, Russian Federation
    \label{JINR}}
\titlefoot{Institut f\"ur Experimentelle Kernphysik,
     Universit\"at Karlsruhe, Postfach 6980, DE-76128 Karlsruhe,
     Germany
    \label{KARLSRUHE}}
\titlefoot{Institute of Nuclear Physics,Ul. Kawiory 26a,
     PL-30055 Krakow, Poland
    \label{KRAKOW1}}
\titlefoot{Faculty of Physics and Nuclear Techniques, University of Mining
     and Metallurgy, PL-30055 Krakow, Poland
    \label{KRAKOW2}}
\titlefoot{Universit\'e de Paris-Sud, Lab. de l'Acc\'el\'erateur
     Lin\'eaire, IN2P3-CNRS, B\^{a}t. 200, FR-91405 Orsay Cedex, France
    \label{LAL}}
\titlefoot{School of Physics and Chemistry, University of Lancaster,
     Lancaster LA1 4YB, UK
    \label{LANCASTER}}
\titlefoot{LIP, IST, FCUL - Av. Elias Garcia, 14-$1^{o}$,
     PT-1000 Lisboa Codex, Portugal
    \label{LIP}}
\titlefoot{Department of Physics, University of Liverpool, P.O.
     Box 147, Liverpool L69 3BX, UK
    \label{LIVERPOOL}}
\titlefoot{Dept. of Physics and Astronomy, Kelvin Building,
     University of Glasgow, Glasgow G12 8QQ
    \label{GLASGOW}}
\titlefoot{LPNHE, IN2P3-CNRS, Univ.~Paris VI et VII, Tour 33 (RdC),
     4 place Jussieu, FR-75252 Paris Cedex 05, France
    \label{LPNHE}}
\titlefoot{Department of Physics, University of Lund,
     S\"olvegatan 14, SE-223 63 Lund, Sweden
    \label{LUND}}
\titlefoot{Universit\'e Claude Bernard de Lyon, IPNL, IN2P3-CNRS,
     FR-69622 Villeurbanne Cedex, France
    \label{LYON}}
\titlefoot{Dipartimento di Fisica, Universit\`a di Milano and INFN-MILANO,
     Via Celoria 16, IT-20133 Milan, Italy
    \label{MILANO}}
\titlefoot{Dipartimento di Fisica, Univ. di Milano-Bicocca and
     INFN-MILANO, Piazza della Scienza 2, IT-20126 Milan, Italy
    \label{MILANO2}}
\titlefoot{IPNP of MFF, Charles Univ., Areal MFF,
     V Holesovickach 2, CZ-180 00, Praha 8, Czech Republic
    \label{NC}}
\titlefoot{NIKHEF, Postbus 41882, NL-1009 DB
     Amsterdam, The Netherlands
    \label{NIKHEF}}
\titlefoot{National Technical University, Physics Department,
     Zografou Campus, GR-15773 Athens, Greece
    \label{NTU-ATHENS}}
\titlefoot{Physics Department, University of Oslo, Blindern,
     NO-0316 Oslo, Norway
    \label{OSLO}}
\titlefoot{Dpto. Fisica, Univ. Oviedo, Avda. Calvo Sotelo
     s/n, ES-33007 Oviedo, Spain
    \label{OVIEDO}}
\titlefoot{Department of Physics, University of Oxford,
     Keble Road, Oxford OX1 3RH, UK
    \label{OXFORD}}
\titlefoot{Dipartimento di Fisica, Universit\`a di Padova and
     INFN, Via Marzolo 8, IT-35131 Padua, Italy
    \label{PADOVA}}
\titlefoot{Rutherford Appleton Laboratory, Chilton, Didcot
     OX11 OQX, UK
    \label{RAL}}
\titlefoot{Dipartimento di Fisica, Universit\`a di Roma II and
     INFN, Tor Vergata, IT-00173 Rome, Italy
    \label{ROMA2}}
\titlefoot{Dipartimento di Fisica, Universit\`a di Roma III and
     INFN, Via della Vasca Navale 84, IT-00146 Rome, Italy
    \label{ROMA3}}
\titlefoot{DAPNIA/Service de Physique des Particules,
     CEA-Saclay, FR-91191 Gif-sur-Yvette Cedex, France
    \label{SACLAY}}
\titlefoot{Instituto de Fisica de Cantabria (CSIC-UC), Avda.
     los Castros s/n, ES-39006 Santander, Spain
    \label{SANTANDER}}
\titlefoot{Inst. for High Energy Physics, Serpukov
     P.O. Box 35, Protvino, (Moscow Region), Russian Federation
    \label{SERPUKHOV}}
\titlefoot{J. Stefan Institute, Jamova 39, SI-1000 Ljubljana, Slovenia
     and Laboratory for Astroparticle Physics,\\
     \indent~~Nova Gorica Polytechnic, Kostanjeviska 16a, SI-5000 Nova Gorica, Slovenia, \\
     \indent~~and Department of Physics, University of Ljubljana,
     SI-1000 Ljubljana, Slovenia
    \label{SLOVENIJA}}
\titlefoot{Fysikum, Stockholm University,
     Box 6730, SE-113 85 Stockholm, Sweden
    \label{STOCKHOLM}}
\titlefoot{Dipartimento di Fisica Sperimentale, Universit\`a di
     Torino and INFN, Via P. Giuria 1, IT-10125 Turin, Italy
    \label{TORINO}}
\titlefoot{INFN,Sezione di Torino, and Dipartimento di Fisica Teorica,
     Universit\`a di Torino, Via P. Giuria 1,\\
     \indent~~IT-10125 Turin, Italy
    \label{TORINOTH}}
\titlefoot{Dipartimento di Fisica, Universit\`a di Trieste and
     INFN, Via A. Valerio 2, IT-34127 Trieste, Italy \\
     \indent~~and Istituto di Fisica, Universit\`a di Udine,
     IT-33100 Udine, Italy
    \label{TU}}
\titlefoot{Univ. Federal do Rio de Janeiro, C.P. 68528
     Cidade Univ., Ilha do Fund\~ao
     BR-21945-970 Rio de Janeiro, Brazil
    \label{UFRJ}}
\titlefoot{Department of Radiation Sciences, University of
     Uppsala, P.O. Box 535, SE-751 21 Uppsala, Sweden
    \label{UPPSALA}}
\titlefoot{IFIC, Valencia-CSIC, and D.F.A.M.N., U. de Valencia,
     Avda. Dr. Moliner 50, ES-46100 Burjassot (Valencia), Spain
    \label{VALENCIA}}
\titlefoot{Institut f\"ur Hochenergiephysik, \"Osterr. Akad.
     d. Wissensch., Nikolsdorfergasse 18, AT-1050 Vienna, Austria
    \label{VIENNA}}
\titlefoot{Inst. Nuclear Studies and University of Warsaw, Ul.
     Hoza 69, PL-00681 Warsaw, Poland
    \label{WARSZAWA}}
\titlefoot{Fachbereich Physik, University of Wuppertal, Postfach
     100 127, DE-42097 Wuppertal, Germany
    \label{WUPPERTAL}}
\addtolength{\textheight}{-10mm}
\addtolength{\footskip}{5mm}
\clearpage
\headsep 30.0pt
\end{titlepage}
%
\pagenumbering{arabic} 
\setcounter{footnote}{0} %
\large

\noindent
\renewcommand{\deg}{^\circ}


\def\Journal#1#2#3#4{{#1}{\bf #2}(#4) #3}
\def\PLB{{ Phys. Lett.}  \bf B}
\def\EUR{{ Eur. Phys. J.} \bf C}
\def\PRL{{ Phys. Rev. Lett.} }
\def\NIMA{{Nucl. Instr. and Meth.} \bf A}
\def\PRD{{ Phys. Rev.} \bf D}
\def\ZPC{{ Zeit. Phys.} \bf C}
\def\CPC{{Comput. Phys. Commun.} }



\section{Introduction}

The cross-section for the doubly resonant production of W bosons has been
measured with the data sample collected with the DELPHI detector
during the high-energy operation phase of the LEP $e^+e^-$ collider (LEP2),
at centre-of-mass energies from 161~GeV up to 209~GeV. Measurements of
total cross-sections, W angular differential distributions
and decay branching fractions, and the value of the CKM element
$|V_{cs}|$ are presented.
They are compared to the Standard Model including most  
recent theoretical predictions~\cite{lep2mcws}.

\begin{figure}[htb]
 \begin{center}
  \epsfig{file=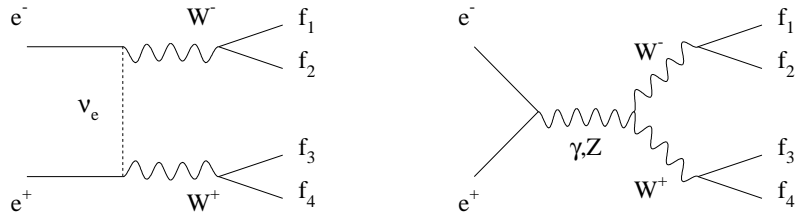,width=12.0cm}
 \end{center}
\vspace{-1.0cm}
\caption{CC03 diagrams.}
\label{fig:cc03} \end{figure}

The cross-sections determined in these analyses correspond
to W~pair production, defined through the three doubly resonant 
tree-level diagrams (``CC03 diagrams''~\cite{CC03} in the following) 
involving $s$-channel $\gamma$ and $Z$ 
exchange and $t$-channel $\nu$ exchange, as shown in Figure~\ref{fig:cc03}. 
Depending on the decay mode of each W, fully hadronic, mixed 
hadronic-leptonic (``semi-leptonic'') or fully leptonic 
final states are obtained.

This paper is organised as follows: 
after a brief description of the DELPHI detector in Section~\ref{sec:detector},
a summary of the analysed data and simulation samples is provided in
Section~\ref{sec:sampl}. Track selection and particle identification
are briefly illustrated in Section~\ref{sec:tracks} and in
Section~\ref{sec:cs} the selection 
of WW events is described and the performance of the analysis 
reported; Section~\ref{sec:syst} is devoted to the discussion of
the systematic error assessment. In 
Sections~\ref{sec:diffxsec} and~\ref{sec:sigtot},
results in terms of differential cross-sections, total cross-section 
and W branching ratios are presented. In Section~\ref{sec:rww} the results
are compared to the theoretical predictions and conclusions follow 
in Section~\ref{sec:conclusions}.

\section {The DELPHI detector in the LEP2 phase}
\label{sec:detector}

The DELPHI detector configuration for LEP1 running was described
elsewhere~\cite{DET}. For operation at LEP2,
changes were made to the subdetectors, the trigger system~\cite{trigger}, 
the run control and the algorithms used in the offline reconstruction
of charged particles, which improved the performance compared to 
LEP1.

The major changes were the extension of the Vertex Detector (VD) and 
the inclusion of the Very Forward Tracker 
(VFT)~\cite{vft}, which enlarged the coverage of the silicon tracker out to 
$11^{\circ} < \theta < 169^{\circ}$\footnote{The DELPHI 
coordinate system has the $z$-axis 
collinear with the incoming electron beam. $\theta$ indicates the polar angle 
with respect to the $z$-axis, $R \phi$ indicates the plane perpendicular to the
$z$ axis.}. 
Also the Inner Detector, both the Jet
Chamber and Trigger Layers, were extended to cover the polar
angle region $15^{\circ} < \theta < 165^{\circ}$.
Together with improved tracking algorithms, alignment 
and calibration procedures optimised for LEP2, these changes led to 
an improved track reconstruction efficiency in the forward regions.

Changes were made to the electronics of the trigger and timing system 
which improved the stability of the running during data taking. The
trigger conditions were optimised for LEP2 running, to give high
efficiency for Standard Model two-fermion and four-fermion processes
and also give  
sensitivity to events which may be signatures of new physics.
In addition, improvements were made to the operation of the detector during
the LEP cycle, to prepare the detector for data taking at the very start
of stable collisions of the beams, and to respond to adverse background
conditions from LEP were they to arise. 
These changes led to an overall improvement
of $\sim10\%$ in the efficiency for collecting the delivered luminosity, 
from $\sim85\%$ at the start of LEP2 in 1995 to $\sim95\%$ at the end in 
year 2000.

During the operation of the detector in year 2000 one of the sectors 
representing 1/12 of the acceptance of the central tracking device, 
the Time Projection Chamber (TPC), failed. This problem affects
about a quarter of the data collected in that year. 
Nevertheless, the redundancy of the tracking system meant that charged 
particles passing through the sector
could still be reconstructed from signals in other tracking 
detectors. A modified track reconstruction algorithm was used in
this sector, which included space points reconstructed in the Barrel RICH
detector, helpful in the determination of the polar angle of charged particles.
As a result, the track reconstruction efficiency was only slightly reduced in 
the region covered by the broken sector. 
The impact of the failure of this part of the detector
on the analyses is discussed further in Section~\ref{sec:systematics}.

\section {Data and simulation samples}
\label{sec:sampl}

A summary of the data samples used for the WW cross-section
measurement is reported in Table~\ref{tab:enlum}, where
the luminosity-weighted centre-of-mass energies and the 
amount of data analysed at each energy are shown.
The luminosity is determined from Bhabha scattering measurements
making use of the very forward electromagnetic calorimetry~\cite{STIC}.
The total integrated luminosity for the LEP2 period corresponds 
approximately to 670~pb$^{-1}$.
The luminosities used for the different selections correspond to those data
for which all elements of the detector essential to each specific analysis
were fully functional; tighter requirements on the detectors used for
lepton identification were applied for the data samples used in the
semi-leptonic and fully leptonic channel analyses.
The luminosity in year 2000 was delivered in a continuum of energies,
thus data taken during this period are divided into two centre-of-mass
energy ranges, above and below 205.5~GeV, referred to as 205~GeV and 
207~GeV in the following.

All the data taken from the year 1997 onwards have been reprocessed with an
updated version of the DELPHI reconstruction software, using a consistent 
treatment for all the samples. 
Larger simulation samples were realised with more up-to-date Monte Carlo 
programs, interfaced to the full DELPHI simulation program DELSIM~\cite{DET} 
and reconstructed with the same reconstruction program as the real data.

The cross-section analyses on these
data are updated with respect to the previously published
ones~\cite{previouspap2_189} and supersede them. 
The data taken in year 1996 have not been reanalysed, because possible
  improvements were negligible compared to the large statistical
  errors of those measurements; these results correspond
to the publications in~\cite{previouspap1}, with a revised determination 
of the luminosity~\cite{previouspap2_183}.

\begin{table}[htb]
\begin{center}
\begin{tabular}{|c|c|c|c|}
\hline
Year & ${\cal L}$-weighted $\sqrt{s}$ (GeV) & Hadronic ${\cal L}$
(pb$^{-1}$)&  Leptonic ${\cal L}$ (pb$^{-1}$) \\
\hline
 1996 &  161.31 &  10.07 &  10.07 \\
      &  172.14 &  10.12 &  10.12  \\
\hline
 1997 &  182.65 &   52.51 &  51.63  \\
\hline
 1998 &  188.63 &  154.35 & 153.81 \\
\hline
 1999 &  191.58 &   25.16 &  24.51  \\
      &  195.51 &   76.08 &  71.99 \\
      &  199.51 &   82.79 &  81.82 \\
      &  201.64 &   40.31 &  39.70 \\
\hline
 2000 &  204.81 &   82.63 &  74.93  \\
      &  206.55 &  135.82 & 123.66 \\
\hline
\end{tabular}
\vspace{0.2cm}
\caption{\label{tab:enlum} Integrated luminosity-weighted centre-of-mass energies 
and luminosities in the LEP2 data taking period. The hadronic luminosity
is used for the four-quark channel, the leptonic one for the semi-leptonic
and fully leptonic channels.}
\end{center}
\end{table}

Four-fermion simulation samples were produced with the
\WPHACT~\cite{wphact} generator, interfaced with the 
\PYTHIA~6.156~\cite{PYTHIA} hadronisation model. In order to perform
checks on hadronisation effects, the same generator was also interfaced
to the \ARIADNE~\cite{ARIADNE} and
\HERWIG~\cite{HERWIG} hadronisation models.

The generation was complemented 
with two-photon collision generators 
BDK~\cite{BDK}, BDKRC~\cite{BDKRC} and \PYTHIA.
The most recent $\mathcal{O}$($\alpha$) electroweak corrections, via
the so-called Leading Pole Approximation (LPA), were included in our generation
of the signal via weights given by the \YFSWW~program~\cite{YFSWW},
according to the scheme described in~\cite{DPA}. 

The selection efficiencies were defined with respect to the CC03 diagrams 
only, by reweighting each event to
the CC03 contribution according to the ratio of the squared matrix elements 
computed with these diagrams only and with the full set of diagrams.

The simulation of two-fermion background processes was realised with the 
KK2f~\cite{KK2f} and \KoralZ~\cite{koralz} generators
interfaced to \PYTHIA, \ARIADNE and \HERWIG, for which the fragmentation
parameters have been tuned at the Z-resonance~\cite{tuning}, and the
\BHWIDE~\cite{bhwide} generator for Bhabha events.

\section{Charged particle selection and lepton identification}
\label{sec:tracks}

To improve the event reconstruction and reject contributions from either
cosmic ray events, beam-gas interactions or noise from
electronics, the reconstructed charged particles were required to fulfil 
the following criteria:

\begin{itemize}
\item[-] momentum 0.1~GeV/c~$<$p$<$~1.5~$\cdot$~P$_{beam}$;
\item[-] momentum error $\Delta$p/p$<$ 1; 
\item[-] $R \phi$ impact parameter $<$ 4~cm;
\item[-] $z$ impact parameter $<$ 4~cm/$\sin\theta$.
\end{itemize}
Tracks seen by only the central tracking devices 
(Vertex Detector (VD) and Inner Detector (ID))
were rejected if no $z$ coordinate measurement was available. 

Neutral clusters of energy in the barrel (HPC), forward (FEMC) and very
forward (STIC) electromagnetic calorimeters were required to have
an energy of at least 300, 400 and 300 MeV respectively. 
In addition, off--momentum electrons in the STIC were rejected by a
cut on the polar angle of a shower at 3$^{\circ}$.
Noise from the hadron calorimeter (HCAL) was reduced by rejecting 
showers which had activity in only one 
layer and failed a selection based on energy, number of hits and longitudinal 
position.

Muon identification was performed by extrapolating tracks through the
entire detector, and associating nearby HCAL energy deposits and
muon chamber hits to the tracks. The errors on track parameters, on the
energy deposits and on the position of 
chamber hits were taken into account when making this
association. Particles were identified as muons if there was at least
one muon chamber hit associated to a track or if the size and
longitudinal profile of the HCAL energy deposits associated to a track
were consistent with a minimum ionizing particle.

Electron identification was based on the reconstructed showers in the 
electromagnetic calorimeters associated to charged particle tracks.
The barrel photons in a conic region close to the direction of the 
track were reassociated to the candidate electron, also accounting for 
the bending caused by the magnetic field. A particle was identified as
an electron if the associated calorimetric energy was larger than 50\%
of the momentum reconstructed in magnetic field. 
For tracks below 30 GeV/c, the identification algorithm was based on a 
combination of the
energy-momentum ratio, the measurement of the energy loss in the TPC,
the matching of the track to the shower in the $z$ and $\phi$
directions and on the shower profile.
FEMC electromagnetic deposits close in space were clustered together and
the association with reconstructed tracks was used for
electron/photon discrimination.
Care was taken in excluding those tracks which were likely to come from the
development of a shower outside the calorimeter.
In addition, the ratio between the electromagnetic energy
and the total energy (electromagnetic and hadronic) was required to be 
above 90\%.
For energetic and isolated tracks in the regions not covered by electromagnetic
calorimetry, the measurement of the energy loss in the TPC was used.
The estimate of electron energy was obtained from a combination of the
track momentum and the calorimetric deposition.

\section{Event selection and partial cross-sections}
\label{sec:cs}

\subsection{Fully hadronic final state}
\label{sec:cs_a}

\subsubsection{Selection of fully hadronic final state events}

For the selection of $\W^{+}\W^{-}\rightarrow 
q\overline{q}q\overline{q}$,
the charged and neutral particles in each event were forced into a
four-jet configuration with the DURHAM 
algorithm~\cite{DURHAM}. 
A preselection of events was performed to select well reconstructed
hadronic events without missing energy and momentum. The following 
conditions were imposed:
\begin{itemize}
\item{reconstructed effective centre-of-mass energy \cite{sprime} 
      $\sqrt{s'} > 65 \%$ of the nominal centre-of-mass energy;}
\item{total and transverse energy for charged particles, $E_{ch} = \sum_{i=1}^{N} E_i^{ch}$
      and $E_t=\sum_{i}{\mid p_{t}^{i} \mid}$, where $p_{t}^{i}$ is the momentum
      component of the particle $i$ perpendicular to the beam axis, 
      each $> 20 \%$ of nominal centre-of-mass energy;}
\item{total particle multiplicity $\geq 3$ for each jet;}
\item{$y_{cut} > 0.0006$ for the migration from 4 to 3 jets 
      when clustering with the DURHAM algorithm;}
\item{convergence of a four-constraint (4C) fit of the 
      measured jet energies and directions imposing four-momentum 
      conservation.}
\end{itemize}

A feed-forward neural network 
was then used to improve the rejection of two-fermion (mainly $\Z/\gamma
\rightarrow q \overline{q}(g)$) and four-fermion background (mainly $\Z \Z
\rightarrow q\bar{q}q\bar{q},q\bar{q}\tau^+\tau^-$). 
The network, based on the \mbox{JETNET} package~\cite{jetnet}, 
uses the standard back-propagation algorithm and consists
of three layers with 13 input nodes, 7 hidden nodes and one output node. 

The following jet and event observables were chosen as input variables,
taking into account previous neural network studies~\cite{somap} 
to optimise input variables for the WW and two-fermion separation:
\begin{itemize}
\item{the difference between the maximum and minimum jet energies after
      the 4C fit;}
\item{the minimum angle between two jets after the 4C fit;}
\item{the value of $y_{cut}$ from the DURHAM algorithm
      for the migration of 4 jets into 3 jets;}
\item{the minimum particle multiplicity of any jet;}
\item{the reconstructed effective centre-of-mass energy $\sqrt{s'}$;}
\item{the maximum probability, amongst the three possible jet
    pairings, for a 6C fit (imposing 4-momentum conservation and
      the invariant mass of each jet pair equal to the W-mass);}
\item{the thrust;}
\item{the sphericity;}
\item{the transverse energy;}
\item{the sum of the cubes of the magnitudes of the momenta of the 7 highest 
      momentum particles $\sum_{i=1}^{7} |\vec{p}_{i}|^{3}$;}
\item{the minimum jet broadening $B_{min}$ \cite{DURHAM};}
\item{the Fox-Wolfram-moment H3 \cite{fox};}
\item{the Fox-Wolfram-moment H4.}
\end{itemize}

\begin{figure}[htb]
 \begin{center}
  \mbox{\epsfig{file=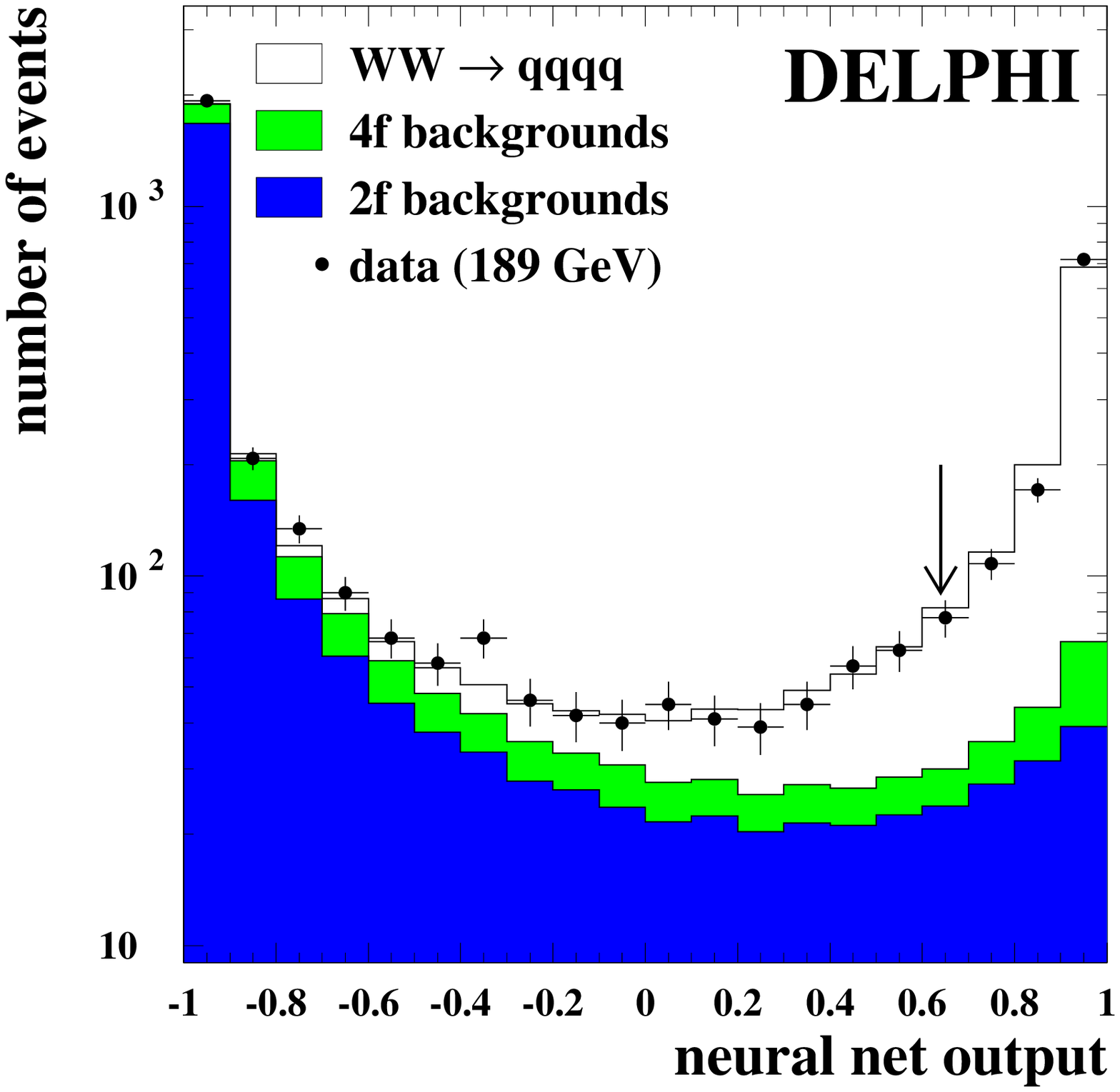,width=8.0cm}
        \epsfig{file=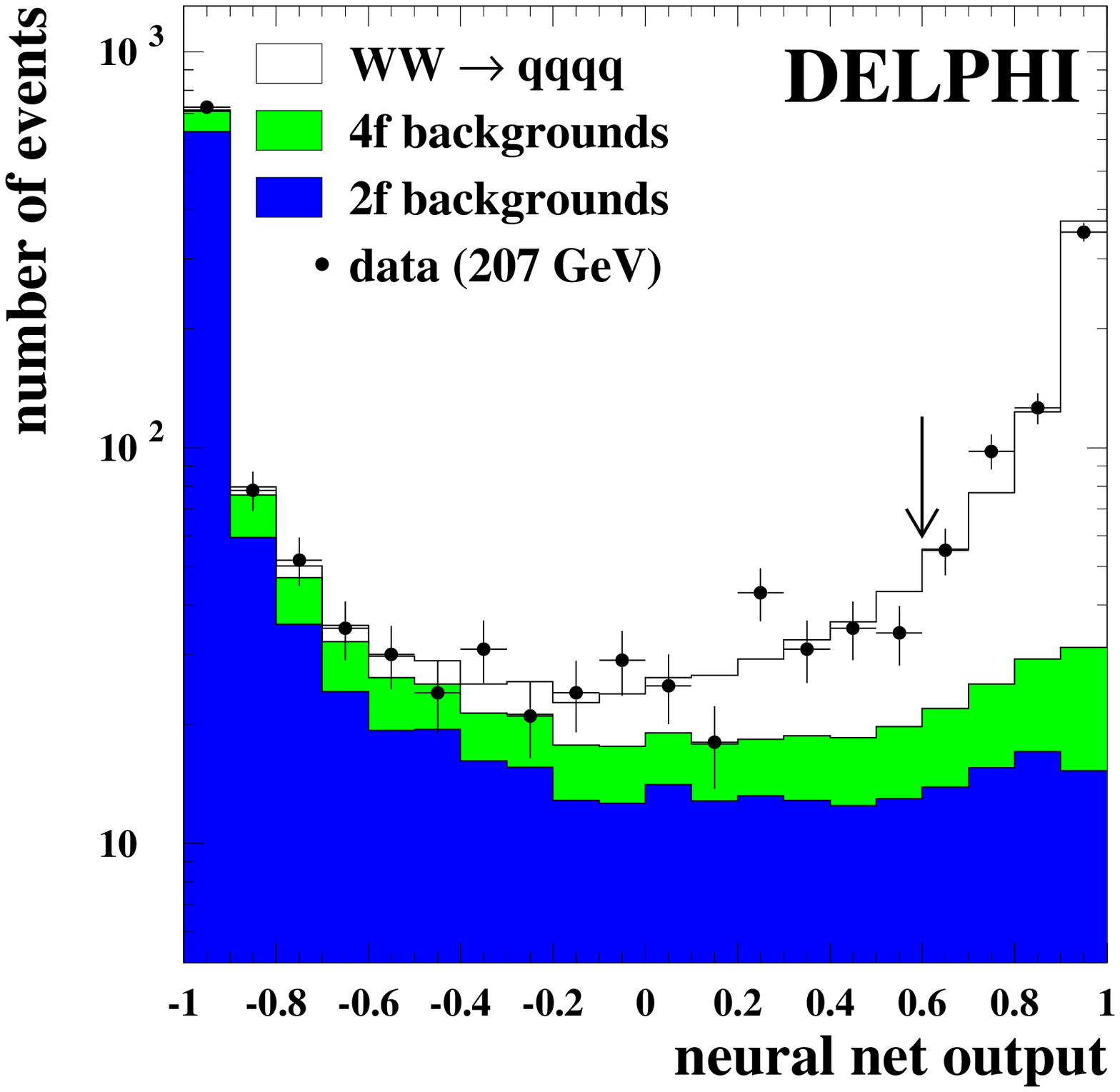,width=8.0cm}}
 \end{center}
\vspace{-1.0cm}
\caption{Distribution of the Neural Network 
output variable for four-jet events
at the centre-of-mass energies of 189~GeV and 207~GeV.
The period of bad TPC operating conditions is excluded
from the plot.
The points show the data and the histograms are the
predicted distributions for signal and background. 
The arrows indicate the cut value applied for the selection of events.}
\label{fig:qqqq} \end{figure}

The neural network was trained separately at each energy. Each training
was performed with 2500 signal events and 2500 $\Z/\gamma\ra\qq $ 
background events. Afterwards
the network output was calculated for other independent samples of simulated
four-fermion, $\Z/\gamma$ and $\gamma\gamma$ events, and for the data.
Figure~\ref{fig:qqqq} shows distributions of the neural network output for
data and simulated events at 189 and 207 GeV.

Events were selected by applying a cut on the NN output variable, chosen 
to minimise the total error on the measured cross-section including the 
systematic uncertainty on the two-fermion background with its correlation at 
all centre-of-mass energies (see Section~\ref{sec:syst}). 

\subsubsection{Results for fully hadronic final state}

The efficiency and background contamination for the
hadronic event selections were evaluated
independently at the different centre-of-mass energies.
The selection performance at $\sqrt{s}$~=~200~GeV
and the total number of events selected in each data sample are 
reported in Table~\ref{qqxx}. 
The efficiencies varied by no more than 4\% over the different energy
points above 172 GeV. The background is dominated by qq$(\gamma)$,
representing 70-75\% of the contamination, decreasing with energy.

\begin{table}[htb]
\begin{center}
\begin{tabular}{|c|c|c|c|c|}
\cline{2-5}
\multicolumn{1}{} {} & \multicolumn{4}{|c|}{efficiencies for selected
channels} \\ 
\hline
channel & $\jjjj$ & $\jjenu$ & $\jjmunu$ & $\jjtaunu$ \\
\hline
$\qq\qq$   & {\bf~0.797~} &   $<10^{-4}$  &   $<10^{-4}$   & 0.012  \\
$\qqenu$   &      0.004   & {\bf ~0.677~} &       0.004    & 0.114  \\
$\qqmunu$  &      0.002   &       0.001   & {\bf ~0.852~}  & 0.043  \\
$\qqtaunu$ &      0.016   &       0.032   &       0.026    & {\bf ~0.581~}\\
\hline
background (pb) & 1.21   & 0.232 & 0.075 & 0.371  \\
\hline
\hline
$\sqrt{s}$ (GeV) & \multicolumn{4}{|c|}{Selected events} \\
\hline
  161         &   15    &   \multicolumn{3}{|c|}{12} \\ 
  172         &   65    &    14      &    17       &    14 \\ 
  183         &  345    &    94      &   118       &   123 \\ 
  189         & 1042    &   269      &   336       &   339 \\ 
  192         &  187    &    42      &    53       &    58 \\ 
  196         &  532    &   151      &   166       &   164 \\ 
  200         &  614    &   162      &   190       &   208 \\ 
  202         &  317    &    94      &    89       &    83 \\ 
  205         &  657    &   169      &   153       &   174 \\
  207         &  999    &   214      &   259       &   289 \\
\hline
  All         & 4773    &   \multicolumn{3}{|c|}{4054} \\ 
\hline
\end{tabular}
\vspace{0.2cm}
\caption{\label{qqxx} Data for the cross-section measurement
of hadronic and semi-leptonic final states. The efficiency 
matrix and the backgrounds are the ones at 200~GeV. The backgrounds
include two-fermion and non-CC03 four-fermion contributions.
The upper limits on the efficiencies are at the 95\% C.L..}
\end{center}
\end{table}

At each energy point the cross-section for fully hadronic events 
was obtained from a binned maximum likelihood fit to the distribution
of the NN output variable above the cut value, assuming Poissonian 
probability density functions for the number of events. The probability
is calculated on the basis of the efficiency  for being reconstructed
in a given bin of the NN output and the expected 
background in each bin and is a function of the partial cross-section
to be measured. 
For this fit the contamination from other WW channels, with its value
fixed to the SM prediction, was considered as a background.

The results for 
$\sigma_{\W\W}^{\qq\qq}$~=~$\sigma_{\W\W}\times$~BR~$(\W\W\rightarrow
\qq\qq)$,  
where BR$(\W\W\rightarrow \qq\qq)$
is the probability for 
the W-pair to give a purely hadronic final state, are reported in
Table~\ref{tab:sigqq}. Systematic uncertainties were determined as
detailed in Section~\ref{sec:syst}.

\begin{table}[htb]
\begin{center}
\begin{tabular}{|c|c|}
\hline
~$\sqrt{s}$ (GeV)~&  ~$\sigma_{\W\W}^{\qq\qq}= \sigma_{\W\W} \times 
   {\mathrm{BR}}(\W\W\rightarrow \qq\qq)$ (pb) \\
\hline \hline
 161 & $1.53^{+0.67}_{-0.55}~\mbox{(stat)} \pm 0.13~\mbox{(syst)} $ \\ 
 172 & $4.65^{+0.95}_{-0.86}~\mbox{(stat)} \pm 0.18~\mbox{(syst)} $ \\ 
 183 & $7.23 \pm 0.45~\mbox{(stat)} \pm 0.09~\mbox{(syst)} $ \\ 
 189 & $7.38 \pm 0.27~\mbox{(stat)} \pm 0.09~\mbox{(syst)} $ \\ 
 192 & $7.78 \pm 0.68~\mbox{(stat)} \pm 0.10~\mbox{(syst)} $ \\ 
 196 & $7.69 \pm 0.39~\mbox{(stat)} \pm 0.10~\mbox{(syst)} $ \\ 
 200 & $7.73 \pm 0.37~\mbox{(stat)} \pm 0.10~\mbox{(syst)} $ \\ 
 202 & $7.83 \pm 0.54~\mbox{(stat)} \pm 0.10~\mbox{(syst)} $ \\ 
 205 & $8.26 \pm 0.38~\mbox{(stat)} \pm 0.10~\mbox{(syst)} $ \\ 
 207 & $7.59 \pm 0.29~\mbox{(stat)} \pm 0.10~\mbox{(syst)} $ \\ \hline
\end{tabular}
\vspace{0.2cm}
\caption{\label{tab:sigqq} Measured hadronic cross-sections. 
}
\end{center}
\end{table}

\subsection{Semi-leptonic final state}   

\subsubsection{Selection of semi-leptonic final state events}
\label{qqlvsel}

Events in which one of the W bosons decays into a lepton and a neutrino
and the other one 
into quarks are characterised by two or more hadronic jets, one isolated lepton 
(coming either directly from the W decay or from the cascade decay 
$\W \rightarrow \tau\nu_{\tau} \rightarrow e\nu_e\nu_{\tau}\nu_{\tau}
\: {\mathrm or} \: \mu\nu_{\mu}\nu_{\tau}\nu_{\tau}$) 
or a low multiplicity jet due to a hadronic
$\tau$ decay, and missing momentum resulting from the neutrino(s).  
The major background comes from $\qqga$ production and from four-fermion 
final states containing two quarks and two leptons of the same flavour.

Events were first required to pass a general hadronic preselection:
\begin{itemize}
\item{at least 5 charged particles;}
\item{energy of charged particles at least 10\% of total centre-of-mass energy;}
\item{$\sqrt{EMF_f^2 + EMF_b^2} < 0.9 \times E_{beam}$, where $EMF_{f,b}$
identify the total energy deposited in electromagnetic calorimeters in 
the forward and backward 
directions, defined as two 20$^{\circ}$ cones around the beam axes.}
\end{itemize}
A search for leptons was then made.
Of the identified electrons with an energy greater than 5 GeV, the one 
with the highest value of (energy $\times \ \theta_{iso}$)
\footnote{The isolation angle $\theta_{iso}$ is defined as the angle made 
to the closest charged particle with a momentum greater than 1 GeV/c.}
was considered to be the electron candidate. 
This candidate was required to have an energy of at least 15 GeV.
Of the identified muons with a momentum greater than 5 GeV/c, the one 
with the highest value of (momentum $\times \ \theta_{iso}$) 
was considered to be the muon candidate. This candidate was required to have a
momentum of at least 15 GeV/c.
The event was then clustered into jets using the LUCLUS~\cite{PYTHIA}
algorithm with a  
$d_{join}$ value of 6.5~GeV/c. The resulting jets were trimmed by removing
particles at an angle greater than 20$^{\circ}$ to the highest energy 
particle in the jet. 
Of these trimmed jets, the one with the smallest momentum--weighted spread
\footnote{defined as $\sum_{i} (\theta_{i} \cdot \mid$$p_{i}$$\mid) /
 \sum_{i} \mid$$p_{i}$$\mid $, where $\theta_{i}$ is the angle made by the 
momentum $p_{i}$ of the $i^{th}$ particle in the jet with the total 
jet momentum}
was taken to be the tau candidate. Particles with momenta smaller than 
1~GeV/c were removed from the candidate jet if they were at an angle 
greater than 8$^{\circ}$ to the jet axis. 
At the end of the procedure, this jet was required to still contain at 
least one charged particle.

An event could have up to three lepton candidates, one of each flavour.

For each lepton candidate, all particles other than the lepton were 
clustered into two jets using the DURHAM algorithm. 
These two jets were required to contain at least three particles, 
at least one of which had to be charged.
Additional preselection cuts are listed in Table \ref{tab:presel_qqlv}:
these reject most events due to photon--photon collisions, 
some events for which there is no missing energy,  
and events whose topologies are far from those of WW events.
\begin{table}[htb]
  \begin{center}
  \begin{tabular}{|r|c|c|}
\hline
                          & $\qqenu$/$\qqmunu$ & $\qqtaunu$ \\
\hline
Transverse energy (GeV)  & $>45$  &  $>40$  \\
Missing momentum (GeV/c) & $>10$  &  10$<$p$_{mis}$$<$80 \\
Visible energy/E$_{train}$ (\%) & 
                             40$<$E/E$_{train}$$<$110 & 
                             35$<$E/E$_{train}$$<$100  \\
Fitted W mass (GeV/c$^2$)  & $>50$  & $>50$ \\
\hline
\end{tabular}
  \end{center}
 \caption{Semi--leptonic preselection cuts. E$_{train}$ indicates
                          the centre-of-mass energy chosen for the
                          training of the IDA (see text). The W mass is fitted through
                          a constrained fit of the measured jets,
                          lepton and unobserved neutrino imposing four-momentum 
                          conservation and the equality of the two W masses in the event.}
 \label{tab:presel_qqlv}
\end{table}

After these preselection cuts, the final selection was made with an
Iterative Discriminant Analysis (IDA) \cite{ida}. The standard IDA 
technique assumes that the signal and background distributions in the 
multi-observable space have different means but identical shapes. IDA was 
extended to treat correctly cases when the distributions have different 
shapes. The input observables were transformed to make their distributions 
Gaussian.
The IDA selection was made in three channels 
($\qqmunu$, $\qqenu$ and $\qqtaunu$).
The training was performed on Monte Carlo samples: around 50k events each of 
four-fermion charged and neutral current processes, and 100k $\qqga$ events. 
The IDA was trained at three centre-of-mass
energies: 189, 200 and 206 GeV.

The observables used in the discriminants are summarised in Table~\ref{tab:vars_qqlv}, 
and can be grouped into four categories:
\begin{itemize}
\item{event observables such as the multiplicity, visible and 
transverse energies, aplanarity\footnote{the aplanarity is 
defined as $\frac{3}{2} \lambda_3$,
where $\lambda_3$ is the smallest eigenvalue of the sphericity tensor
$ S^{\alpha \beta} = \frac{\sum_i p^{\alpha}_{i} p^{\beta}_{i} } 
{\sum_i |\mathbf{p}_i|^2 } $. The $p_i$ are the 3-momenta of particles in 
the event, and $\alpha, \beta = 1,2,3$ correspond to the $x, y, z$ 
momentum components.}
and the reconstructed effective centre-of-mass energy 
($\sqrt{s'}$) are useful to distinguish between semi-leptonic and other
four-fermion events, and to remove $\qqga$ events, particularly those in the
radiative return peak;}
\item{observables concerned with the charged lepton 
(energy, isolation angle, tau jet multiplicity) are useful in discriminating
between events with primary leptons and those with leptons originating
in quark decays or in other processes, such as photon conversion;}
\item{properties of the missing momentum (magnitude and polar angle) make 
use of the fact that the missing momentum from unseen initial state radiation
(ISR in what follows) photons is directed along the 
beampipe;}
\item{the angles between the lepton and missing momentum, 
and the fitted W mass from a constrained kinematic fit 
(imposing four-momentum conservation and the equality of the two W masses 
in the event) are sensitive to the event topologies expected from semi-leptonic 
WW decays.}
\end{itemize}

\begin{table}[htb]
  \begin{center}
  \begin{tabular}{|l|c|c|c|}
\hline
                                 & $\qqenu$ & $\qqmunu$ & $\qqtaunu$ \\
\hline
total multiplicity               & X  & X  &  X \\
visible energy                   & X  & X  &  X \\
transverse energy                & X  &    &    \\
aplanarity                       &    &    &  X \\
$\sqrt{s'}$/$\sqrt{s}$           & X  & X  &  X \\
lepton energy                    & X  & X  &    \\
lepton isolation angle           & X  & X  &  X \\
tau jet charged multiplicity     &    &    &  X \\
tau jet momentum--weighted spread&    &    &  X \\
magnitude of missing momentum    & X  & X  &  X \\
polar angle of missing momentum  & X  & X  &  X \\
lepton--missing momentum angle   &    & X  &  X \\
Fitted W mass                    & X  & X  &  X \\
\hline
\end{tabular}
  \end{center}
 \caption{Observables in semi--leptonic channels used in IDA analysis.}
 \label{tab:vars_qqlv}
\end{table}

The performance of the selection was measured using samples
independent from those on which the training was performed.
Events were selected with a cut on the output of the IDA, chosen to optimise
the product of efficiency and purity for each channel. 
Events were first passed to the $\qqmunu$ selection; if they were not selected,
they were passed to the $\qqenu$; if they were still not selected, 
they were then finally passed to the $\qqtaunu$ selection.

\begin{figure}[htb]
 \begin{center}
  \begin{tabular}{ccc}
  \hspace{-1cm} 
  \includegraphics[width=5.4cm,height=7cm]{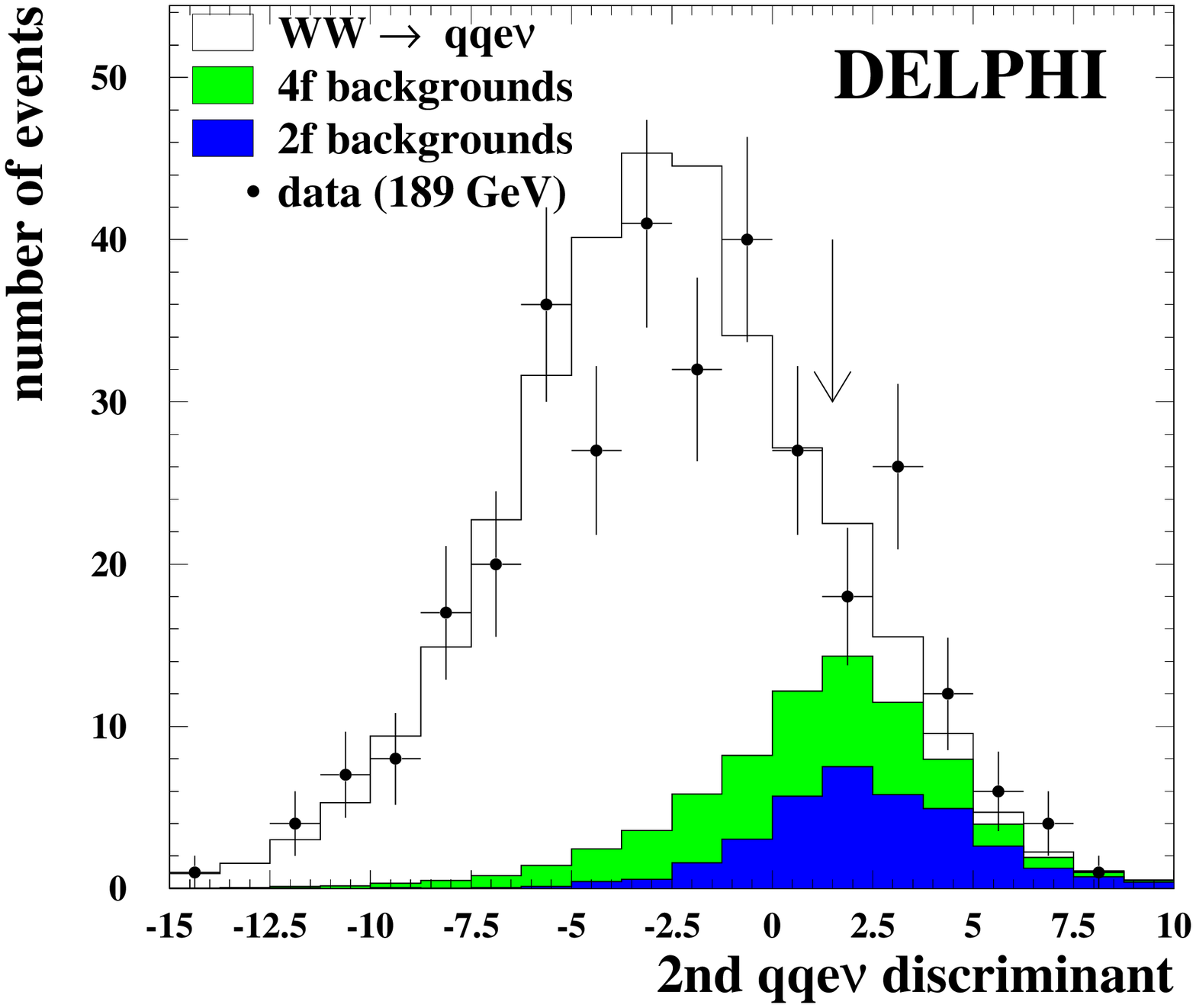} & 
  \includegraphics[width=5.4cm,height=7cm]{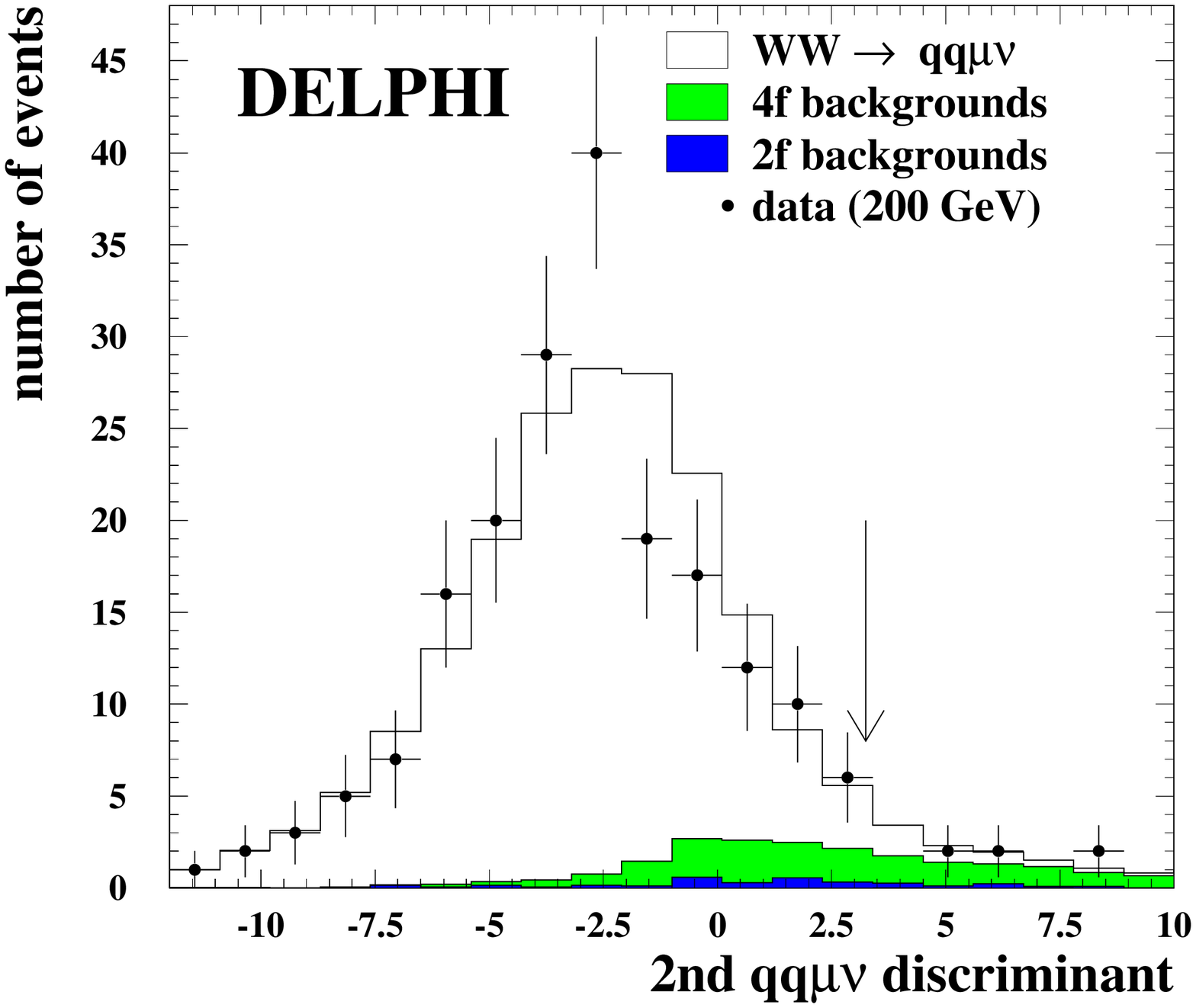} & 
  \includegraphics[width=5.4cm,height=7cm]{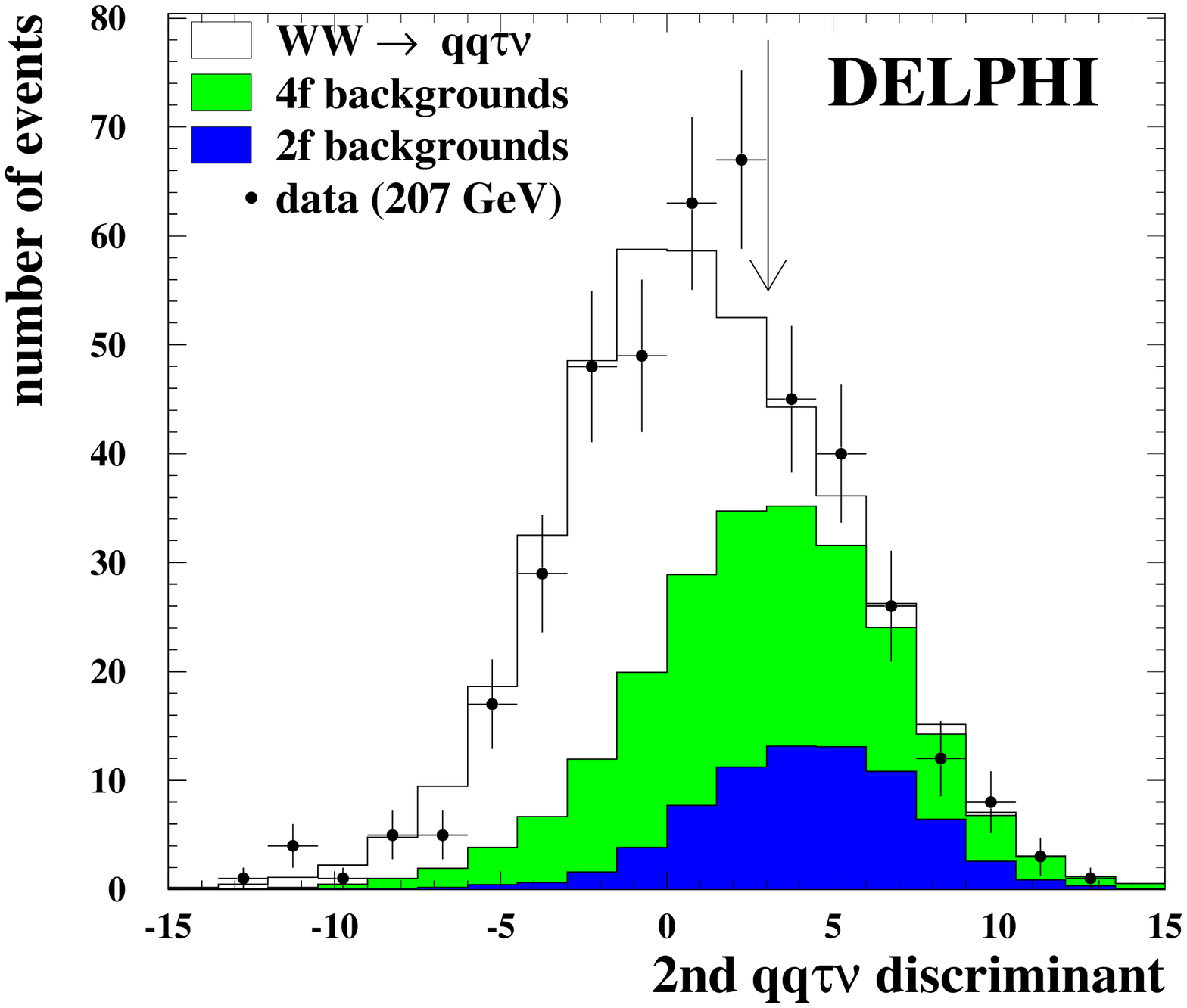} \\
  \end{tabular}
  \vspace{-1.cm} 
 \end{center}
 \caption{Distribution of discriminants for the semi-leptonic
   selection. The plots refer to the $\qqenu$ selection at 189~GeV,
   the $\qqmunu$ selection at 200~GeV
   and the $\qqtaunu$ selection at 207~GeV. The arrows indicate the
   cut value applied for the selection of events. The points show the data 
   and the histograms are the predicted distributions for signal and background.}
 \label{fig:qqlv}
\end{figure}

Distributions of discriminants for the semi-leptonic event
selection are shown in Figure~\ref{fig:qqlv}.

\subsubsection{Results for semi-leptonic final state}

The efficiency matrices and background contaminations for the
semi-leptonic event selections were evaluated
independently at the different centre-of-mass energies.
The efficiencies differed by no more than 2\% over the different energy 
points.
The values at 200 GeV, with the number of events observed at each 
energy point are reported in Table~\ref{qqxx}. 
The final efficiencies for identifying $\qqenu$, $\qqmunu$ and 
$\qqtaunu$ events with any of the three semi-leptonic selections at
200 GeV were 79.6\%, 89.6\% and 63.9\% respectively.
The background in all the channels was composed 
of two- and four-fermion events in similar proportions.

The number of events observed in the different lepton channels was
found to be consistent with lepton universality.
With this assumption, likelihood fits to the Poisson probability of
the expected number of events, where the contamination of the fully hadronic 
channel is considered as background,
yield the cross-sections $\sigma_{\W\W}^{qql\nu}= \sigma_{\W\W} \times 
{\mathrm{BR}}(\W\W\rightarrow \qq l\nu)$ reported in Table~\ref{tab:sigql}.
Systematic uncertainties were determined as
detailed in Section~\ref{sec:syst}.

\begin{table}[htb]
\begin{center}
\begin{tabular}{|c|c|}
\hline
~$\sqrt{s}$ (GeV)~&  ~$\sigma_{\W\W}^{qql\nu}= \sigma_{\W\W} \times 
   {\mathrm{BR}}(\W\W\rightarrow \qq l\nu)$ (pb) \\
\hline \hline
 161 & $1.74^{+0.67}_{-0.55}~\mbox{(stat)} \pm 0.10~\mbox{(syst)} $ \\ 
 172 & $5.68^{+1.02}_{-0.93}~\mbox{(stat)} \pm 0.17~\mbox{(syst)} $ \\ 
 183 & $7.25 \pm 0.46~\mbox{(stat)} \pm 0.08~\mbox{(syst)} $ \\ 
 189 & $6.89 \pm 0.26~\mbox{(stat)} \pm 0.08~\mbox{(syst)} $ \\ 
 192 & $6.88 \pm 0.64~\mbox{(stat)} \pm 0.08~\mbox{(syst)} $ \\ 
 196 & $7.50 \pm 0.39~\mbox{(stat)} \pm 0.08~\mbox{(syst)} $ \\ 
 200 & $7.82 \pm 0.37~\mbox{(stat)} \pm 0.08~\mbox{(syst)} $ \\ 
 202 & $7.71 \pm 0.53~\mbox{(stat)} \pm 0.08~\mbox{(syst)} $ \\ 
 205 & $7.45 \pm 0.38~\mbox{(stat)} \pm 0.08~\mbox{(syst)} $ \\ 
 207 & $6.95 \pm 0.29~\mbox{(stat)} \pm 0.08~\mbox{(syst)} $ \\ \hline
\end{tabular}
\vspace{0.2cm}
\caption{
\label{tab:sigql} Measured semi-leptonic cross-sections. 
}
\end{center}
\end{table}

\subsection{Fully leptonic final state}

\subsubsection{Selection of fully leptonic final state event}

Events in which both W bosons decay into $l\nu$ are characterised by low
multiplicity, a clean two-particle (or jet from $\tau$ decay) topology 
with two energetic, acollinear leptons of opposite charge, acoplanar with
the beam and with large missing momentum and 
energy. The relevant backgrounds are di-leptons from 
$e^+e^- \rightarrow \Z (\gamma)$, Bhabha scattering, two-photon
collisions and leptonic final states from Z-pair and single boson production.

The selection was performed in three steps. First a leptonic
preselection was made, followed by particle identification. 
Finally two Neural Networks were used to reject 
the remaining background.

An initial set of cuts was applied to select a sample enriched in leptonic
events. All particles in the event were clustered into 
``jets'' using the LUCLUS 
algorithm ($d_{join}=6.5~{\mathrm GeV}/c$) 
and only events with two
reconstructed jets, containing at least one charged particle each, were
retained. A total charged particle multiplicity 
between 2 and 6 was required and at least one jet had to have
only one charged particle. In order to reduce the background from 
two-photon collisions and radiative di-lepton events,
the event acoplanarity, $\theta_{acop}$, defined as the acollinearity of the 
two jet directions projected onto the plane perpendicular to the beam axis, 
had to be above 3$^{\circ}$. 
In addition, the total momentum transverse to the beam direction, $P_t$,
had to exceed 2\% of the centre-of-mass energy $\sqrt{s}$.
The associated energy deposited in the electromagnetic calorimeters
for both leading particles (the ones with the largest momenta) was 
required to be less than $0.44\cdot \sqrt{s}$ to reject Bhabha
scattering. 
To reject radiative events further, the energy of the most energetic 
photon had to be less than $0.25\cdot\sqrt{s}$ and the angle in the
plane perpendicular to the beam axis 
between the charged particles system and the most energetic photon was required
to be less than 170$^{\circ}$. Finally, the energy of the charged particles
in each
jet had to be greater than $0.04\cdot\sqrt{s}$ and the visible energy 
of the particles with $\mid \cos\theta\mid <0.9$ had to exceed 
$0.06\cdot\sqrt{s}$.

In events passing this selection each particle was classed as 
$\mu$, $e$ or hadron. 
 A lepton was identified as a cascade decay 
from $\W \rightarrow \tau \nu_\tau$ if the momentum was lower 
than $0.13\cdot\sqrt{s}$.

After the preselection and the channel identification, two Neural Networks 
based on the Multi Layer Perceptron (MLP)~\cite{MLP} package
were built to reject the remaining background.
They consisted of one output layer from 13 input variables, normalised 
to lie in the region between zero and one. 

One of the two Neural Networks was tuned for the
$\tau\nu_\tau X\nu_X$ channels ($X=e,\mu,\tau$) 
and the other for the remaining channels, 
given the different characteristics of the two samples. This was found
to optimise the performance of the selection. 

\begin{figure}[htb]
 \begin{center}
  \mbox{\epsfig{file=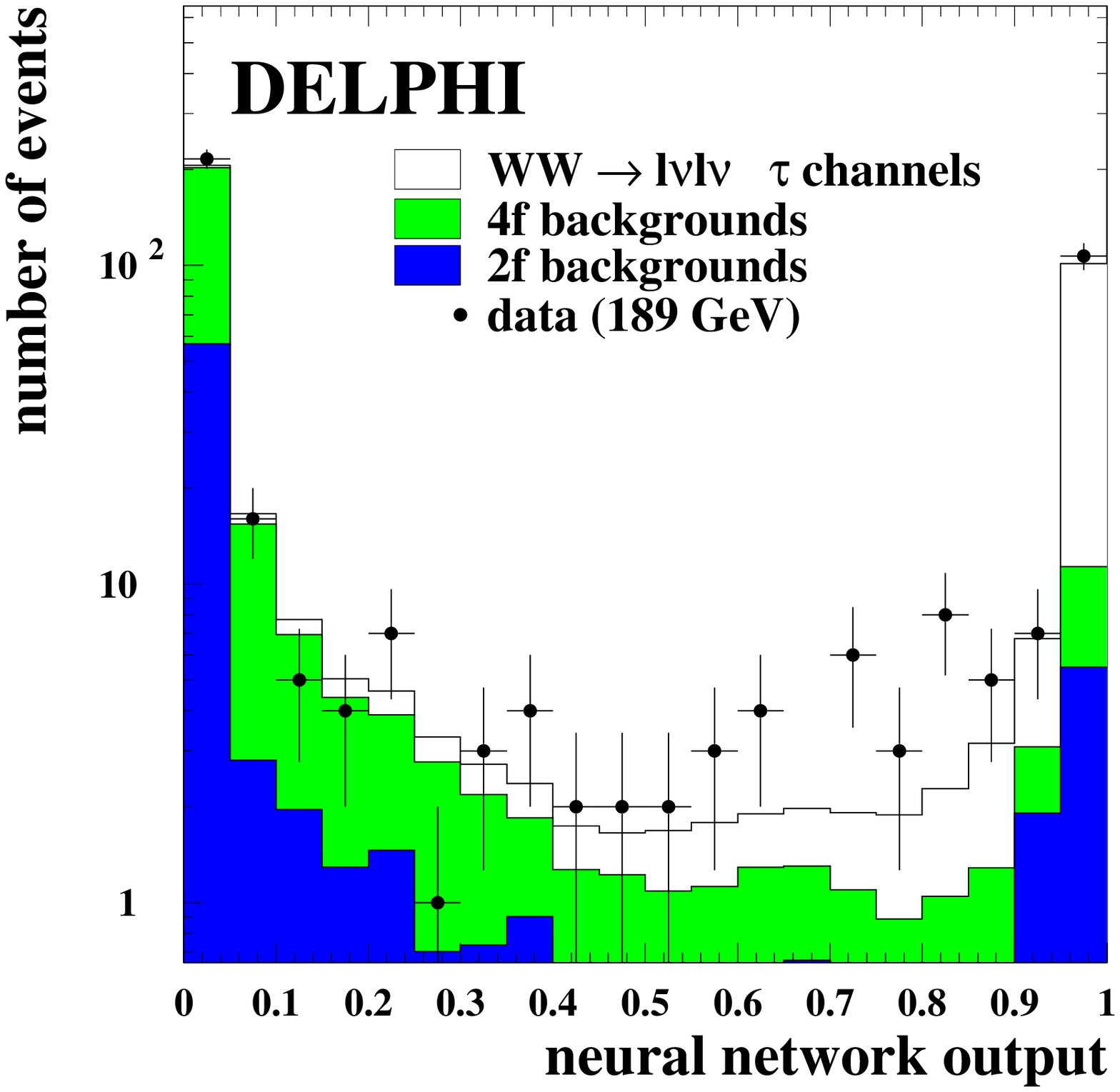,width=8.0cm}
        \epsfig{file=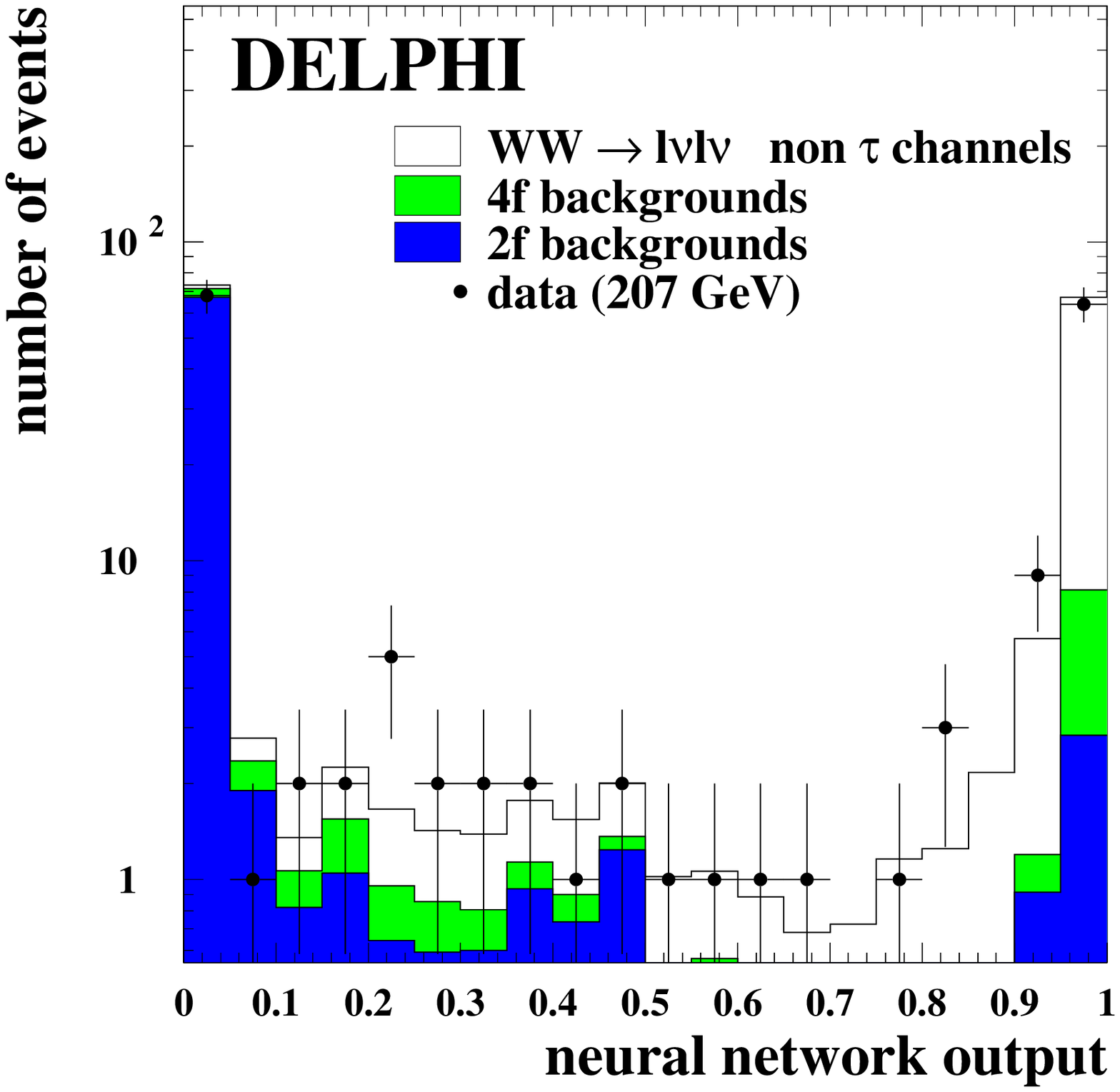,width=8.0cm}}
 \end{center}
\vspace{-1.0cm}
\caption{
Distribution of the output variable of the two types of Neural Network
used for the selection of  fully-leptonic events.
The points show the data and the histograms are the
predicted distributions for signal and background. }
\label{fig:lvlv} \end{figure}

The following variables were used in both Neural Networks:

\begin{itemize}
\item
the event acoplanarity;
\item
the event acollinearity;
\item
the larger of the associated energy from electromagnetic calorimetry
of the two leptons;
\item
the transverse momentum, $p_t$;
\item
the transverse energy;
\item
the angle in the plane perpendicular to the beam axis
between the vector sum of the charged particle momenta and the most
energetic photon; 
\item
the absolute value of the cosine of the polar angle of the missing momentum;
\item
the energy in the calorimeters not associated to charged particles 
outside two cones of 20$^{\circ}$ around both leading charged particles;
\item
the larger of the energies of the two jets coming from charged particles;
\item
the total energy of neutral particles;
\item
the larger of the invariant masses of the two jets;
\item
the total visible energy;
\item
the total energy of charged particles.
\end{itemize}
Distributions of the Neural Network output variable 
for the fully leptonic event
selection are shown in Figure~\ref{fig:lvlv}.
The cut applied on the Neural Network output was tuned in order to
optimise the product of efficiency and purity. The cut was channel
dependent, but did not depend on the centre-of-mass energy.

\subsubsection{Results for fully leptonic final state}

The efficiencies and backgrounds for 
$\sqrt{s} = 200$ GeV together with the numbers of selected events 
at each of the centre-of-mass energies are shown in Table~\ref{lnulnu}. 
The overall $l\nu l\nu$ efficiency was 67.3\% and the residual 
background from non-W and single-W events was 0.187~pb. 
The efficiencies were checked to be constant within 1-2\%
at the different energy points above 172 GeV.
The number of observed events in each subchannel is consistent with the
hypothesis of lepton universality.
Table~\ref{tab:sigll} presents the values, at each energy, of 
$\sigma_{\W\W}^{l\nu l\nu}= \sigma_{\W\W} \times 
{\mathrm{BR}}(\W\W\rightarrow l\nu l\nu)$ from maximum likelihood fits
to the Poisson probability of the expected number of events.
Systematic uncertainties were determined as
detailed in Section~\ref{sec:syst}.

\begin{table}[htb]
\begin{center}
\begin{tabular}{|c|c|c|c|c|c|c|}
\cline{2-7}
\multicolumn{1}{} {} & \multicolumn{6}{|c|}{efficiencies for selected
channels} \\ 
\hline
channel & $\tau\nu\tau\nu$ & $e\nu\tau\nu$ & $\mu\nu\tau\nu$ & $e\nu e\nu$
& $e\nu\mu\nu$ & $\mu\nu\mu\nu$ \\ \hline
$\tau\nu\tau\nu$ & {\bf 0.272} & 0.087 & 0.077 & 0.004 & 0.008 & 0.006 \\ 
$e\nu\tau\nu$    &   0.068 & {\bf 0.462} & 0.005 & 0.044 & 0.049 & $<2\cdot
10^{-3}$ \\ 
$\mu\nu\tau\nu$  &   0.042 & 0.004 & {\bf 0.536} & $<2\cdot 10^{-3}$ & 0.061
& 0.053 \\ 
$e\nu e\nu$      &   0.017 & 0.151 & $<10^{-3}$ & {\bf 0.471} & $<10^{-3}$ &
$<2\cdot 10^{-3}$ \\
$e\nu\mu\nu$     &   0.012 & 0.041 & 0.094 & $<2\cdot 10^{-3}$ &{\bf 0.621}
& $<10^{-3}$ \\ 
$\mu\nu\mu\nu$   &   0.008 & $<10^{-3}$ & 0.109 & $<2\cdot 10^{-3}$ &
0.003 & {\bf 0.677} \\ 
\hline
background (pb)  &   0.030 & 0.046 & 0.023 & 0.042 & 0.019 & 0.027 \\
\hline 
\hline 
$\sqrt{s}$ (GeV) & \multicolumn{6}{|c|}{Selected events} \\
\hline
161 &    \multicolumn{6}{|c|}{2}    \\
172 &    \multicolumn{6}{|c|}{8}    \\
183 &     4   &  13   &   10  &    9  &    9   & 14    \\
189 &    22   &  56   &   43  &   25  &   45   & 38    \\
192 &     4   &   2   &   14  &    4  &    8   &  7    \\
196 &    16   &  29   &   19  &   12  &   24   &  7    \\
200 &    12   &  26   &   28  &   11  &   27   & 13    \\
202 &     4   &  20   &   13  &    4  &    6   &  9    \\
205 &    14   &  26   &   22  &    7  &   24   & 10    \\
207 &    14   &  40   &   41  &   16  &   44   & 16    \\
\hline
All &    \multicolumn{6}{|c|}{891}    \\
\hline
\end{tabular}
\vspace{0.2cm}
\caption{\label{lnulnu} Data for the cross-section measurement
of the fully leptonic final state. The efficiency 
matrix and the background are the ones at 200~GeV. The upper limits on
the efficiencies are at 95\% C.L.}
\end{center}
\end{table}

\begin{table}[htb]
\begin{center}
\begin{tabular}{|c|c|}
\hline
~$\sqrt{s}$ (GeV)~&  ~$\sigma_{\W\W}^{l\nu l\nu}= \sigma_{\W\W} \times 
   {\mathrm{BR}}(\W\W\rightarrow l\nu l\nu)$ (pb) \\
\hline \hline
 161 & $0.30^{+0.39}_{-0.24}~\mbox{(stat)} \pm 0.09~\mbox{(syst)} $ \\ 
 172 & $1.03^{+0.50}_{-0.39}~\mbox{(stat)} \pm 0.09~\mbox{(syst)} $ \\ 
 183 & $1.59 \pm 0.26~\mbox{(stat)} \pm 0.08~\mbox{(syst)} $ \\ 
 189 & $1.86 \pm 0.14~\mbox{(stat)} \pm 0.04~\mbox{(syst)} $ \\ 
 192 & $1.97 \pm 0.37~\mbox{(stat)} \pm 0.06~\mbox{(syst)} $ \\ 
 196 & $1.87 \pm 0.21~\mbox{(stat)} \pm 0.05~\mbox{(syst)} $ \\ 
 200 & $1.84 \pm 0.20~\mbox{(stat)} \pm 0.05~\mbox{(syst)} $ \\ 
 202 & $1.81 \pm 0.28~\mbox{(stat)} \pm 0.05~\mbox{(syst)} $ \\ 
 205 & $1.82 \pm 0.20~\mbox{(stat)} \pm 0.06~\mbox{(syst)} $ \\ 
 207 & $1.82 \pm 0.16~\mbox{(stat)} \pm 0.06~\mbox{(syst)} $ \\ \hline
\end{tabular}
\vspace{0.2cm}
\caption{\label{tab:sigll} Measured fully leptonic cross-sections. 
}
\end{center}
\end{table}

\section{Systematic errors}
\label{sec:syst}

A large variety of systematic effects were taken into
account in the cross-section determination. 
They were due to imperfect modelling of the detector response or of 
underlying physics in the simulation, or to
statistical uncertainties due to the finite size of the simulation samples.

Systematic errors were divided in three classes  
to facilitate their treatment in the combination of cross-section
results from the four LEP experiments~\cite{lepcomb}: 
errors correlated between the experiments and between the
different centre-of-mass energies (LCEC), typically due to the use of the
same models to describe physics effects, errors uncorrelated between the
experiments but correlated between centre-of-mass energies (LUEC), 
which comprise detector-related effects, and errors
correlated neither between experiments nor between energies (LUEU) 
which are mainly due to uncertainties on simulation sample statistics.
Correlations between the different channels were also taken into account.

Details of the determination of the systematic contributions are
given in the following.

\subsection{Estimation of systematic uncertainties}
\label{sec:systematics}

\subsubsection{Background cross-sections from four-fermion and two-fermion processes} 

Theoretical uncertainties in the knowledge of four-fermion 
cross-sections largely depended on the process, varying
from $\pm 2\%$ in ZZ processes
to $\pm 5\%$ in single-boson contributions. Larger uncertainties arose when
considering regions of the phase space dominated by
$\gamma\gamma$ collisions~\cite{lepcomb}. 

The four-fermion background generation was performed 
independently for charged and neutral current processes.
The latter ones were further divided with phase space cuts to isolate
as much as possible the contribution of multiperipheral diagrams, 
where $\gamma\gamma$ scattering dominates, and which were generated
in a complementary sample~\cite{DPA}.
For the selected background from charged current processes, mainly the single-W  
contribution, a relative uncertainty of $\pm 5\%$ was assigned.
The same uncertainty was used for the non-$\gamma\gamma$ neutral current
background, with the exception of ZZ contributions where it
became $\pm 2\%$. For the remaining neutral current phase space a
$\pm 10\%$ relative uncertainty was assigned.

The theory errors used for the two-fermion background cross-sections
were those reported in~\cite{lep2f}. A further uncertainty of $\pm 1\%$
was conservatively added to account for ungenerated regions of the phase
space in the matching between two-fermion and four-fermion processes, where 
radiative corrections to $ee\rightarrow f\bar{f}$ could convert into an extra 
low-mass fermion pair.

The variations of each background were taken as fully correlated in channel
and energy and the resulting change in the measured cross-section was 
quoted as a systematic error.

\subsubsection{Modelling of four-jet background from $q\bar{q}$}

In the fully hadronic channel the uncertainty on the background from
two-fermion production leading to four-jet final states was estimated
by comparing 
simulations with different hadronisation models, \ARIADNE, \PYTHIA~and \HERWIG.
\ARIADNE~was chosen as default, because it provided the best description
of the four-jet rates observed for the large data set at the
Z peak~\cite{tuning} and also at LEP2 energies~\cite{Flagmeyer}. For the
other models differences of $(-6.0~\pm~0.4)\%$ (\PYTHIA) and
$(+3.9~\pm~0.6)\%$ (\HERWIG) were obtained for the background
computation, and the largest difference of 6\% was conservatively chosen
as the systematic uncertainty.
This estimate was confirmed by fitting all data
to signal and background with a free parameter to scale the two-fermion 
background (\ARIADNE), for which a change of $(4~\pm~3)\%$ was obtained.

\subsubsection{Fragmentation modelling}

Modelling of the fragmentation in hadronic events could have an impact
both on the selection efficiency and on the background level estimated
from the Monte Carlo. These effects were evaluated by comparing the
performance of the selection algorithms on signal and background 
samples generated with different hadronisation models. 

The \PYTHIA~hadronisation model, which best described the two-jet 
fragmentation at the Z peak, was taken as a reference to evaluate
efficiencies for all channels and the background for the semi-leptonic 
channel. 
It was compared to \ARIADNE~\cite{ARIADNE} and \HERWIG~\cite{HERWIG} and
the largest deviations were considered to estimate 
the systematic errors on the cross-sections. 

The largest variation in the signal efficiency was 0.58\% in the
fully hadronic channel and 1\% in the semi-leptonic one, 
while the difference in the background level was found to be 3.2\% 
in the semi-leptonic channel. For the hadronic channel this uncertainty is
included in the four-jet modelling.

\subsubsection{Final state interactions}

At LEP2 energies the decay distance between the W bosons was 
smaller than the hadronisation scale or the typical radius where 
Bose-Einstein effects occur.
Therefore gluon exchange between quarks from different Ws 
(known as Colour Reconnection) or 
Bose-Einstein Correlations between pions were to be expected.
These so-called Final State Interactions (FSI) between the decay
products of the two different Ws could affect the reconstruction
of fully hadronic events only, and their modelling could have an impact 
on the determination of the selection efficiency.
 
The effects of Colour Reconnections and Bose-Einstein Correlations were
estimated by evaluating the selection efficiency on simulation samples
where FSI were modelled using the SK1
algorithm~\cite{SK1} with reconnection probability of 30\% for 
Colour Reconnections and the LUBOEI algorithm~\cite{BEC} 
for Bose-Einstein Correlations. The 
full difference between the presence and the absence of these effects 
was taken as an indication of the
systematic error, corresponding to a variation in efficiency of 
-0.3\% and 0.2\%, respectively.

\subsubsection{Radiative corrections}

The correct simulation of radiation in WW production and decay could be 
relevant for efficiency determination.
Since the LEP2 Monte Carlo workshop~\cite{lep2mcws}, generators with 
a more precise 
calculation of $\mathcal{O}$($\alpha$) electroweak corrections to 
CC03 became available~\cite{RacoonWW,YFSWW}. The theoretical uncertainty
on the total cross-section was reduced by almost a factor 4 down to 
a level of 0.5\%, with a change of central value by almost 2\% with respect
to Gentle2.0~\cite{Gentle}, run with parameter settings as described 
in~\cite{lep2yr}. Moreover, it was shown that the more correct description
of electroweak corrections  
also had very important effects on differential 
distributions~\cite{chierici}, making its inclusion essential for
efficiency determinations.

The \YFSWW~\cite{YFSWW} program was used in the generation
for reweighting events according to the procedure described 
in~\cite{DPA}. In order to estimate the effect that 
theoretical uncertainty in the description of the radiation had on the 
selection efficiency, the results obtained in the DELPHI setup were
compared with a simulation making use of \RacoonWW~\cite{RacoonWW} 
on the testbench 
process $u\bar{d}\mu\nu_{\mu}$.
The two programs differed in many respects in the treatment of QED 
radiation: \YFSWW~was a $e^+e^-\rightarrow \W^+\W^- \rightarrow 4f$
generator (CC03 diagrams only) with $\mathcal{O}$($\alpha$) 
factorisable electroweak corrections and non-factorisable corrections 
implemented via the so-called Khoze-Chapovsky ansatz (KC)~\cite{KCansatz}.
It included ISR in leading logarithm approximation (LL) 
$\mathcal{O}$($\alpha^3$) via YFS exponentiation~\cite{YFSexp} and
final state radiation (FSR) LL  
$\mathcal{O}$($\alpha^2$) via PHOTOS~\cite{PHOTOS}.
On the other hand, \RacoonWW~was a $4f$ generator implementing 
$\mathcal{O}$($\alpha$) in Double Pole Approximation (DPA) rigorously, extended to 
$\mathcal{O}$($\alpha^3$) for collinear ISR via structure functions.
It also included real corrections with the exact 
$e^+e^-\rightarrow 4f\gamma$ matrix elements of the CC11~\cite{CC03} class.

The systematic uncertainty on the cross-section was estimated
by comparing results from the two programs: no significant difference 
was found in the efficiencies, therefore the statistical error on
this difference was taken as a conservative estimate of the 
systematic contribution.

\subsubsection{Luminosity determination}

The luminosity was determined from a measurement of Bhabha scattering, which
was theoretically known to high accuracy. The measurement made use
of coincidences in the very forward electromagnetic calorimeters
and was affected by the experimental error on the acceptance ($\pm 0.5\%$). 
The residual theoretical uncertainty on the cross-section estimate was
$\pm 0.12\%$~\cite{bhabha_pol}. 
These uncertainties were propagated to a systematic error on $\sigma_{\W\W}$. 

The statistical error on the Bhabha cross-section was included in the 
statistical error of the W-pair cross-section.

\subsubsection{Detector effects}

A non-perfect reproduction of track reconstruction and lepton
identification efficiencies in the simulation could induce 
a systematic error on the signal efficiency and the background rejection. 
This was
particularly relevant for semi-leptonic and fully leptonic channels.

These effects were evaluated from a comparison of the simulation with
data on high statistics samples of clean two-lepton and two-jet
 events, which were collected at a centre-of-mass energy of 91.2~GeV.
The data were taken with the same detector and trigger 
configuration and analysed with the same reconstruction software 
as the high energy data. High energy data samples were also used for
 the comparison, selecting events with a clean di-lepton and 2- or
 3-jet topologies. 

From data-simulation comparisons, corrections were deduced for the
energies and 
polar angles of jets, and for the number of charged particle
tracks in the endcap region. 
These led to corrections of the signal efficiency in the
hadronic channel of $\pm 0.14\%$ (jets) and $\pm 0.30\%$ (endcap
tracks), for which systematic 
uncertainties of $\pm 0.14\%$ and $\pm 0.10\%$ were assigned.

For lepton identification, the angular averaged difference in the 
identification efficiency between data and simulation
of 0.3\% for muons and 1\% for electrons was found.
The systematic error on the WW cross-section and W branching
fractions was determined by 
randomly changing the lepton identification in the signal and
background Monte Carlo according to the discrepancies 
found. The changes were correlated between the different channels.
The effect on the total cross-section was small because most of the 
events lost as electrons or muons were recovered by the tau selections. 

Other possible discrepancies between data and Monte Carlo in the variables used as 
inputs for the Neural Network in the hadronic channel were taken into 
account by smearing these variables in the simulation by their experimental
resolution. 

In total a systematic uncertainty of $\pm 0.55\%$ on the signal efficiency 
and $\pm 1.7\%$ on the total background was estimated for the hadronic channel. 
For the semi-leptonic channels the uncertainties on the signal efficiencies
were $\pm 0.5\%$ for the $qqe\nu$ and $qq\mu\nu$ channels, and $\pm
1\%$ for the  
$qq\tau\nu$. The variations of the background ranged from 0.6\% to 6\%
according to the channel.
The systematic uncertainties in the fully leptonic channel were $\pm 1.5\%$ for
the efficiency and 7\% for the background.

\subsubsection{Detector inefficiencies in a specific period} 

 As mentioned in Section~\ref{sec:detector}, one of the TPC sectors, 
 covering 1/12 of the acceptance, 
 was not operational during the last period of  
 the high energy data taking in the year 2000. 
 These data were analysed separately and then combined with the results 
 from the previous period.
 The performance of the analyses and the cross-section values were
 found to be compatible within statistical errors.
 Additional systematic effects were estimated by comparing 
 data collected at the $\Z$ peak during the period with the TPC sector 6 off
 with simulation samples produced with the same detector
 conditions. Both hadronic and leptonic $\Z$ decays were used. 
 The impact on the WW cross-section analysis was conservatively
 evaluated as an
 uncertainty on the selection efficiency of $\pm 0.5\%$ in the fully hadronic
 channel and of 1\% in the other channels, which was added to the
 systematic error. No effects were found on the background level.

\subsubsection{Monte Carlo statistics}

The uncertainties due to limited statistics of the Monte Carlo samples 
were at most  $\pm 0.2\%$ on the signal efficiency and $\pm 2.0\%$ on the
background level for the hadronic channels at energies above 172 GeV.
The error on backgrounds in the fully leptonic channels were up to 10 times 
larger because of the small numbers of accepted events.

\begin{table}[htb]\centering
\begin{tabular}{|l|c|c|c|cc|}
\hline
Source & $\sigma_{\W\W}^{\qq\qq}$ (pb) & $\sigma_{\W\W}^{\qqlnu}$ (pb)
& $\sigma_{\W\W}^{\lnulnu}$ (pb) & LC & EC \\ 
\hline
Four-jet modelling        & $\pm 0.051$ & $\pm 0.014$ &   -   &   Y  &  Y \\
Background cross-sections & $\pm 0.009$ & $\pm 0.016$ & $\pm 0.006$ &
   Y  &  Y \\
Fragmentation             & $\pm 0.045$ & $\pm 0.038$ &   -   &   Y  &  Y \\
Final state interactions  & $\pm 0.025$ &   -         &   -   &   Y  &  Y \\
Radiative corrections     & $\pm 0.008$ & $\pm 0.008$ & $\pm 0.002$ &
   Y  &  Y \\
Luminosity (theor)        & $\pm 0.011$ & $\pm 0.010$ & $\pm 0.002$ &
   Y  &  Y \\
Luminosity (exp)          & $\pm 0.045$ & $\pm 0.043$ & $\pm 0.011$ &
      &  Y \\
Detector effects          & $\pm 0.045$ & $\pm 0.053$ & $\pm 0.033$ &
      &  Y \\
Monte Carlo statistics    & $\pm 0.005$ & $\pm 0.014$ & $\pm 0.033$ &
      &    \\
\hline
\end{tabular}
\caption{\label{tab:syst} Breakdown of systematic errors on the partial
  $WW$ cross-sections
  at $\sqrt{s}$=200 GeV and their classification according to 
  correlations between LEP experiments (LC) and between centre-of-mass
  energies (EC).}
\end{table}

Table~\ref{tab:syst} presents the breakdown of the systematics per
channel determined in the way described above for 200~GeV. 
The classification of the systematics according to their correlations 
in energy or experiments is also shown.

\section{Differential cross-section}
\label{sec:diffxsec}
Given the high statistics available at LEP2, it is interesting 
to provide a measurement of differential cross-sections.
Particularly relevant are the W polar angle distributions, from which 
triple gauge coupling limits can be extracted.
In what follows the determination of
d$\sigma_{\W\W}$/dcos$\theta_{\W^-}$ is 
discussed. 

The presented differential cross-section refer to CC03 $q\bar{q}e\nu$ and 
$q\bar{q}\mu\nu$ final states, since they give a clean W charge assignment,
and has therefore to be understood as
d[$\sigma_{\W\W}$(BR$_{e\nu}$+BR$_{\mu\nu}$)]/dcos$\theta_{\W^-}$, where
BR$_{l\nu}$ is the branching ratio of the decay WW~$\rightarrow$~$\qqlnu$.

The IDA analysis described in section~\ref{qqlvsel} was used for the event
selection; the W flight direction was determined via a constrained fit 
of the event which imposed four-momentum conservation and equality of 
the hadronic and leptonic invariant masses. In order to distinguish between
radiation from Ws or from the final state fermions, the photon-to-fermion 
recombination scheme followed the CALO5 definition adopted in the LEP2
Monte Carlo workshop~\cite{lep2mcws}. 
To match the detector acceptance best, an additional restriction
requiring the charged lepton to be 
more than 20$^{\circ}$ away from the beam direction 
was introduced. For the $q\bar{q}e\nu$ case this additional cut strongly
suppressed the contribution of single-W diagrams in the signal
definition.
The impact of possible systematic effects due to charge
misidentification for leptons within the accepted angular region
was found to be negligible from studies on dilepton events.
To optimise the statistics in each bin, four bins in energy and ten bins 
in polar angle were chosen. 
Table~\ref{tab:enbin} reports the energy binning, together with the 
corresponding luminosity and luminosity weighted centre-of-mass energy.
The angular binning was chosen
to be significantly bigger than the resolution on $\cos\theta_{\W^-}$, estimated to 
be about 0.06 from Monte Carlo, in order to minimise bin migration of data. 
The migration matrix, expressing the probability that an event selected
in a certain bin was generated in another one, is reported in Appendix~A.
No corrections for bin migration were applied to the presented results.  
All the conventions used for presenting the DELPHI result, including signal 
and bin definitions, are the ones agreed for the LEP combination. 

\begin{table}[htb]
\begin{center}
\begin{tabular}{|c|c|c|}
\hline
Bin (GeV)& ${\cal L}$-weighted $\sqrt{s}$ (GeV) & ${\cal L}$ (pb$^{-1}$) \\
\hline
 180-184 &  182.65  &  51.63  \\
\hline
 184-194 &  189.03  &  178.32  \\
\hline
 194-204 &  198.46 &  193.52  \\
\hline
 204-210 &  205.91  &  198.59  \\
\hline
\end{tabular}
\vspace{0.2cm}
\caption{\label{tab:enbin} Energies and luminosities in the bins for
  the differential cross-section measurement.}
\end{center}
\end{table}

\begin{figure}[htb]
 \begin{center}
  \includegraphics[width=0.9\textwidth]{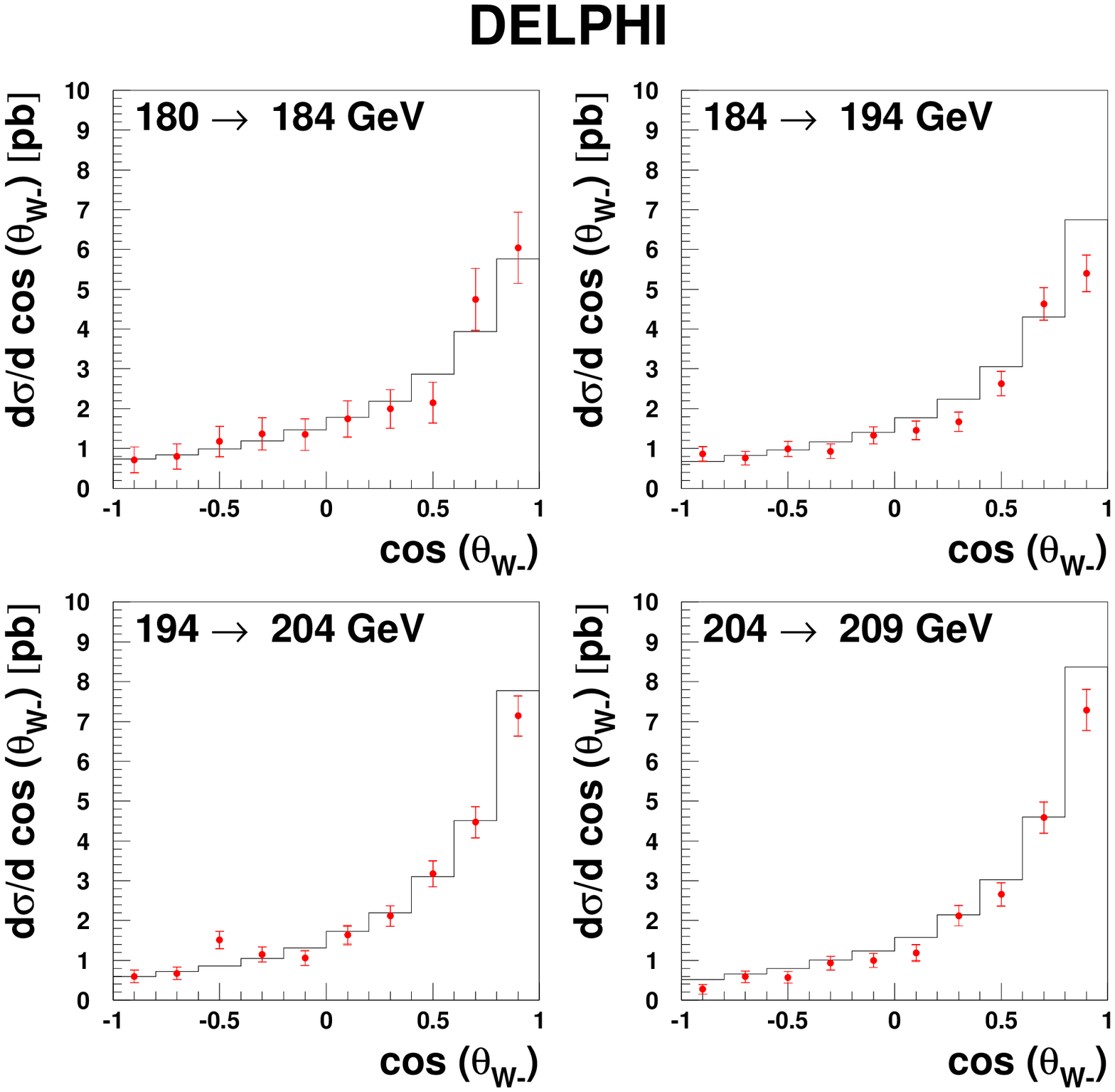}
 \end{center}
\vspace{-1.0cm}
\caption{W$^-$ polar angle differential cross-sections from the $q\bar{q}e\nu$ 
and $q\bar{q}\mu\nu$ channels (detailed definition in the text). The measurements 
in the four energy bins defined in the text (points) are compared with
the expectations (histograms). The systematic contributions are
included in the error bars.} 
\label{dsdcost} 
\end{figure}

The results at the energies in Table~\ref{tab:enbin} 
are reported in Figure~\ref{dsdcost}, where data
points are superimposed on the expected distributions from \WPHACT~and \YFSWW.  
The data are in agreement with the expectations in all energy ranges.
A systematic deficit of data in the highest cos$\theta_{\W^-}$
bin at energies above 184~GeV is observed. 
As a crosscheck, it was verified that the shape of the angular
distributions, for electrons and muons separately, and for positively
and negatively charged leptons, were compatible within errors.

The detailed list of results, in terms of measured cross-sections,
statistical and systematic errors per bin is reported in 
Appendix~A.

\section{Determination of the W branching fractions and of the total 
WW production cross-section}
\label{sec:sigtot}

The total cross-section for WW production and the W leptonic branching 
fractions were obtained from a likelihood fit based
on the probabilities of finding the observed number of events in each final
state. The input numbers in the form given in Tables~\ref{qqxx} 
and~\ref{lnulnu} were used, except for the fully hadronic final state
where, for energies above 172~GeV, the binned distribution of the 
neural network output was used. 

\begin{figure}[htb]
 \begin{center}
  {\epsfig{file=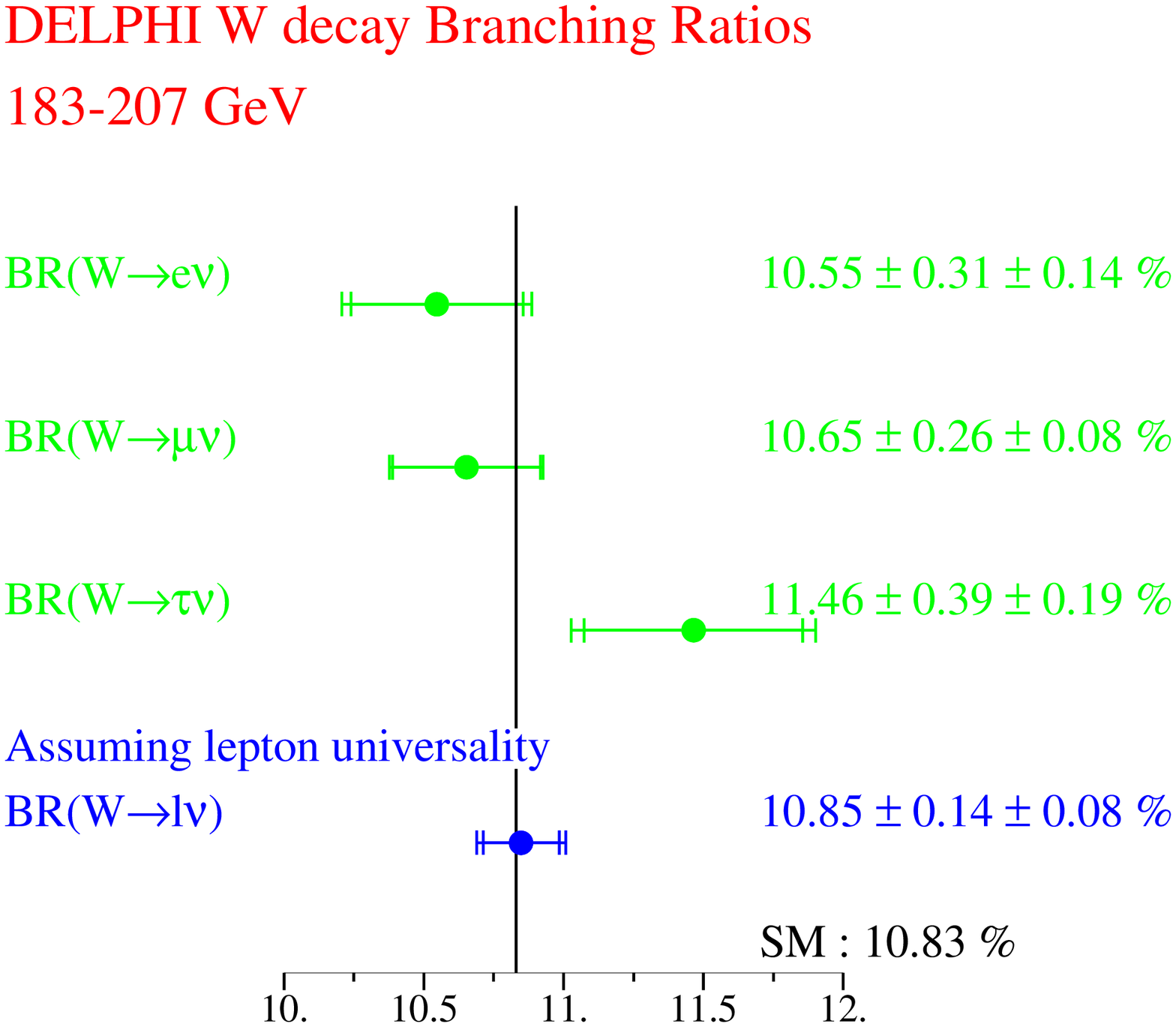,width=12.0cm}}
 \end{center}
\vspace{-1.0cm}
\caption{W decay branching ratios (without and with the assumption of lepton
  universality) measured with the DELPHI data, in
  comparison with the Standard Model expectations~\cite{pdg}. 
  Both the statistical 
  and the total errors are indicated in the error bars.}
\label{fig:br} \end{figure}

A fit without the assumption of lepton universality, requiring the unitarity 
of the branching ratios 
(i.e. BR$_{e\nu}$~+~BR$_{\mu\nu}$~+~BR$_{\tau\nu}$~+~BR$_{q\bar{q'}}$~=~1),
was performed. The results for all the data above 172~GeV are shown in 
Table~\ref{brlept}. The analyses on the low statistics samples at
161 and 172 GeV used inclusive lepton identification.
The correlation matrix is also reported, where both statistical and
systematic contributions are included. 
Since the lepton branching ratios are in agreement, a second fit assuming
lepton universality was performed in order to extract BR$_{q\bar{q'}}$.
The result for this second fit are also reported in Table~\ref{brlept}.
In Figure~\ref{fig:br} the results are compared to the Standard Model
predictions. All results are consistent with the expectations and with lepton universality.  

\begin{table}[htb]
\begin{center}
\begin{tabular}{|l|c|c|cc|}
\hline
channel & branching fraction & stat. error & syst. error (LU) 
& syst. error (LC) \\
\hline
$W \rightarrow e\nu$       &  0.1055 & 0.0031 & 0.0013 & 0.0005 \\
$W \rightarrow \mu\nu$     &  0.1065 & 0.0026 & 0.0006 & 0.0005 \\
$W \rightarrow \tau\nu$    &  0.1146 & 0.0039 & 0.0017 & 0.0009 \\
\hline
\end{tabular}
\vspace{0.3cm}
\begin{tabular}{|l|r r r|}
\hline
Correlations& $\W \rightarrow e\nu$ & $W \rightarrow \mu\nu$ & $W \rightarrow \tau\nu$ \\
\hline
$\W \rightarrow e\nu$    & 1.00 & 0.03 &-0.34 \\
$\W \rightarrow \mu\nu$  & 0.03 & 1.00 &-0.17 \\
$\W \rightarrow \tau\nu$ &-0.34 &-0.17 & 1.00 \\
\hline
\end{tabular}
\vspace{0.7cm}
\begin{tabular}{|l|c|c|cc|}
\cline{2-5}
\multicolumn{1}{} {} & \multicolumn{4}{|c|}{assuming lepton universality} \\
\hline
channel & branching fraction & stat. error & syst. error (LU) 
& syst. error (LC) \\
\hline
$\W \rightarrow {\mathrm{hadrons}}$      
        &  0.6745 & 0.0041 & 0.0020 & 0.0014 \\
\hline
\end{tabular}
\vspace{0.2cm}
\caption{\label{brlept} W branching fractions from 
data above 172~GeV 
and correlation matrix for the leptonic branching fractions.}
\end{center}
\end{table}

Within the Standard Model, 
the branching fractions of the W boson depend on the six matrix elements 
$|\mathrm{V}_{\mathrm{qq'}}|$ of the Cabibbo--Kobayashi--Maskawa (CKM) 
quark mixing matrix not involving the top quark. 
In terms of these matrix elements, 
the leptonic branching fraction of the W boson BR$_{l\nu}$ is given by
$$
  \frac{1}{BR_{l\nu}} \quad  =  \quad 3 
  \Bigg\{ 1 + 
          \bigg[ 1 + \frac{\alpha_s(\mathrm{M}^2_{\mathrm{W}})}{\pi} 
          \bigg] 
          \sum_{\tiny\begin{array}{c}i=(u,c),\\j=(d,s,b)\\\end{array}}
          |\mathrm{V}_{ij}|^2 
  \Bigg\},
$$
where $\alpha_s(\mathrm{M}^2_{\W})$ is the strong coupling
constant. Taking $\alpha_s(\mathrm{M}^2_{\W})=0.119\pm0.002$~\cite{pdg},
the measured leptonic branching fraction of the W yields
$$
  \sum_{\tiny\begin{array}{c}i=(u,c),\\j=(d,s,b)\\\end{array}}
  |\mathrm{V}_{ij}|^2 
  \quad = \quad
  1.996\,\pm\,0.043\,(BR_{l\nu})\,\pm\,0.002\,(\alpha_s),
$$
where the first error 
is due to the uncertainty on the branching fraction measurement 
and the second to the uncertainty on $\alpha_s$. 
Using the experimental knowledge~\cite{pdg} of the sum
$|\mathrm{V}_{ud}|^2+|\mathrm{V}_{us}|^2+|\mathrm{V}_{ub}|^2+
 |\mathrm{V}_{cd}|^2+|\mathrm{V}_{cb}|^2=1.0476\pm0.0074$, 
the results of Table~\ref{brlept} 
can be interpreted as a measurement of $|\mathrm{V}_{cs}|$ 
which is the least well determined of these matrix elements:
$$  |\mathrm{V}_{cs}|\quad=\quad0.973\,\pm\,0.019 {\mbox{(stat)}}
    \,\pm\,0.012 {\mbox{(syst)}}  $$
where the uncertainties on the SM parameters are included 
in the systematic error.

If the SM values of the W branching ratio~\cite{pdg} 
are assumed, the fitting 
procedure can be repeated with the total WW production cross-sections 
as the only free parameter. The results obtained from all the data samples
are reported in Table~\ref{tab:sigww}. In this table the breakdown of the 
systematic contributions into the correlation categories defined in
Table~\ref{tab:syst} is also shown. Correlations between the
different channels were also taken into account.

\begin{table}[htb]
\begin{center}
\begin{tabular}{|c|c|ccc|}
\hline
 $\sqrt{s}$ (GeV)~& $\sigma_{\W\W}$ (pb)~& 
                   \multicolumn{3}{c|} {$\delta_\sigma$ (syst.) (pb)} \\
&  & LCEC & LUEU & LUEC \\
\hline \hline
 161 &~$3.61^{+0.97}_{-0.85}~\mbox{(stat)} \pm 0.19~\mbox{(syst)} $
& 0.039 & 0.182 & 0.037 \\ 
 172 & $11.37^{+1.44}_{-1.35}~\mbox{(stat)} \pm 0.32~\mbox{(syst)} $ 
& 0.115 & 0.288 & 0.095 \\ 
 183 & $16.07 \pm 0.68~\mbox{(stat)} \pm 0.16~\mbox{(syst)} $ 
& 0.100 & 0.044 & 0.126 \\ 
 189 & $16.09 \pm 0.39~\mbox{(stat)} \pm 0.16~\mbox{(syst)} $ 
& 0.094 & 0.028 & 0.127 \\ 
 192 & $16.64 \pm 0.99~\mbox{(stat)} \pm 0.17~\mbox{(syst)} $ 
& 0.103 & 0.040 & 0.131 \\ 
 196 & $17.04 \pm 0.58~\mbox{(stat)} \pm 0.17~\mbox{(syst)} $ 
& 0.100 & 0.033 & 0.133 \\ 
 200 & $17.39 \pm 0.55~\mbox{(stat)} \pm 0.17~\mbox{(syst)} $ 
& 0.102 & 0.029 & 0.135 \\ 
 202 & $17.37 \pm 0.79~\mbox{(stat)} \pm 0.17~\mbox{(syst)} $ 
& 0.107 & 0.030 & 0.136 \\ 
 205 & $17.56 \pm 0.57~\mbox{(stat)} \pm 0.17~\mbox{(syst)} $ 
& 0.107 & 0.039 & 0.136 \\ 
 207 & $16.35 \pm 0.44~\mbox{(stat)} \pm 0.17~\mbox{(syst)} $ 
& 0.103 & 0.051 & 0.128 \\ \hline
\end{tabular}
\vspace{0.2cm}
\caption{\label{tab:sigww} Measured total WW cross-sections. 
  A breakdown of the systematic uncertainties in the different
  contributions, correlated between LEP experiments and/or between
  different centre-of-mass energies, is shown, according to
  the classification described in the text (see
  Section~\ref{sec:syst} for details).}
\end{center}
\end{table}

\begin{figure}[htb]
\centerline{\epsfig{file=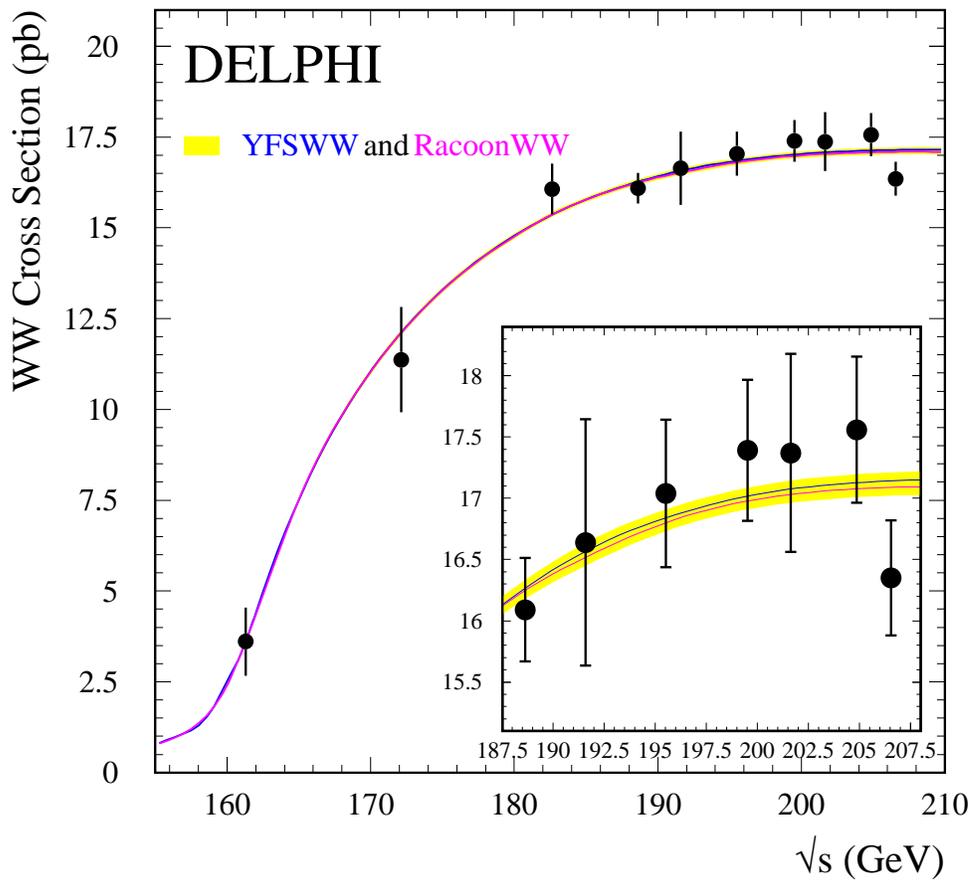,width=14cm}}
\caption{Measurements of the WW cross-section 
compared with the Standard Model prediction given by the \YFSWW~\cite{YFSWW}
and \RacoonWW~\cite{RacoonWW} programs. The shaded band represents the 
uncertainty on the theoretical calculations.}
\label{fig:sigww} \end{figure}

A comparison of the results with the most recent calculations in the Double Pole 
Approximation from \RacoonWW~and \YFSWW~is shown in Figure~\ref{fig:sigww}.
As DPA computations are not reliable close to the WW threshold,
the predictions below 168~GeV were obtained running those programs in the 
Improved Born Approximation (IBA), which only accounts for 
initial state radiation and Coulomb corrections.
The shaded region represents the theoretical uncertainty of the 
calculations and was obtained by an analytic parametrisation of the 
relative uncertainty given in~\cite{lep2mcws}. This led to an accuracy 
on the theoretical curves of about $\pm 0.7\%$ at 168~GeV and of $\pm 0.4\%$ 
at 200~GeV.
The uncertainties from \RacoonWW~and from \YFSWW~were merged into 
a single error band.

The measurements are in very good 
agreement with the Standard Model expectations.

\section{Determination of $\cal{R}_{\W\W}$ }
\label{sec:rww}

The cross-section measurements for different energies can be 
combined into a single value to quantify the overall agreement with 
theoretical predictions. $\rww$ is defined as the ratio between the 
experimentally determined cross-section and the theoretical expectations.
This procedure was used to compare the measurements
at the eight energies between 183 and 207 GeV to the predictions 
of \Gentle, \KoralW, \YFSWW~and \RacoonWW. The measurements at 161 and
172 GeV were not used because of the high sensitivity of the
cross-section to the value of the W mass at these energies.

For each calculation, the cross-sections were converted into
ratios by dividing them by the corresponding theoretical predictions, 
and combined taking into account the energy correlation of the
systematics.

The theoretical errors on the calculations, due to both physical and 
technical precision of the generators used, were also propagated to the 
ratios.

\begin{table}[htb]
\begin{center}
\begin{tabular}{|l|c|}
\hline
Theoretical prediction  &  $\rww$  \\
\hline
~\Gentle 2.0 
& ~0.974 $\pm$ 0.015 (exp) $\pm$ 0.019(theo)~ \\ 
~\KoralW     
& ~0.979 $\pm$ 0.015 (exp) $\pm$ 0.010(theo)~ \\ 
~\YFSWW     
& ~0.999 $\pm$ 0.015 (exp) $\pm$ 0.005(theo)~ \\ 
~\RacoonWW  
& ~1.001 $\pm$ 0.015 (exp) $\pm$ 0.005(theo)~ \\
\hline
\end{tabular}
\vspace{0.2cm}
\caption{\label{tab:rww} $\rww$ values from the combination of all the data,
 using different theoretical calculations.}
\end{center}
\end{table}

\begin{figure}[htb]
 \begin{center}
 \epsfig{file=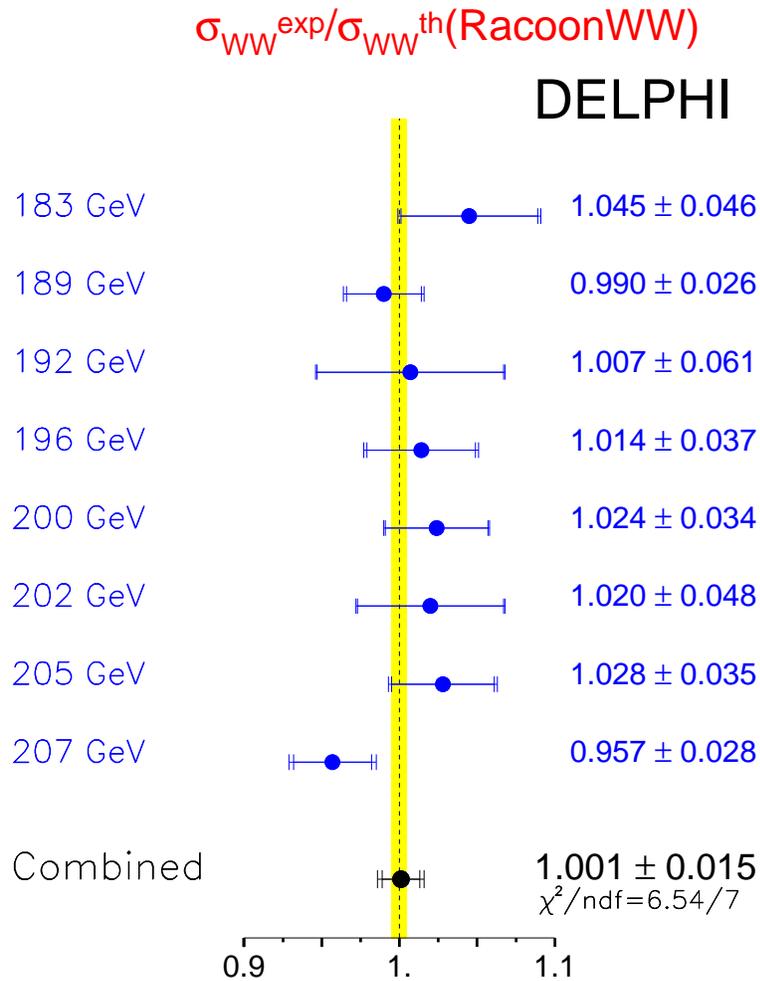,width=13.0cm}
 \end{center}
\vspace{-1.0cm}
\caption{Ratios between measured and predicted CC03 cross-sections with the 
DELPHI data. The error bars indicate the combined statistical and
 systematic uncertainties. The 
energy-combined values are indicated at the bottom. The plot refers
to the \RacoonWW~predictions, where a 0.5\% theory error is indicated
as a band.}
\label{fig_rww} \end{figure}

The values of $\rww$ at the various centre-of-mass energies are
presented in Figure~\ref{fig_rww} for the \RacoonWW~calculation. 
The band on the figure represents the theoretical error on the prediction,
where its dependence on energy is ignored for simplicity.
The $\rww$ values from the combination of all the data
using different theoretical calculations are shown in Table~\ref{tab:rww}. 
Both statistical and systematic contributions are indicated.
It is worth noting how the estimated theory error 
(taken from~\cite{lep2mcws}, page 34) decreases from \Gentle~to \RacoonWW~and 
that the final experimental precision on $\rww$ with 
the DELPHI data at LEP2 approaches $\pm 1.5\%$, 
to be compared with $\pm 0.5\%$ from the best theory computations.
The data favour the more complete inclusion of radiative corrections in
the calculation at the level of 1.5~$\sigma$.

\section{Conclusions}
\label{sec:conclusions}

The WW production cross-section from $e^+e^-$ annihilations has been 
measured at ten centre-of-mass energies between 161 and 209 GeV with 
the DELPHI experiment at LEP. The data correspond to a total integrated 
luminosity of about 670 pb$^{-1}$. 

The results are in agreement with the predictions of the most recent
CC03 cross-section calculations and test them to an accuracy 
of about 1.5\%.

Differential distributions in the polar angle of the reconstructed
W are also measured in the semileptonic channels.

The W branching fractions are measured with an uncertainty of less 
than $\pm 4\%$ for individual leptons and about $\pm 0.7\%$ for hadrons.
They also agree with the Standard Model expectation. From the
leptonic branching fraction a precise determination of $|V_{cs}|$ is
derived.

\subsection*{Acknowledgements}
\vskip 3 mm
 We are greatly indebted to our technical 
collaborators, to the members of the CERN-SL Division for the excellent 
performance of the LEP collider, and to the funding agencies for their
support in building and operating the DELPHI detector.\\
We acknowledge in particular the support of \\
Austrian Federal Ministry of Education, Science and Culture,
GZ 616.364/2-III/2a/98, \\
FNRS--FWO, Flanders Institute to encourage scientific and technological 
research in the industry (IWT), Federal Office for Scientific, Technical
and Cultural affairs (OSTC), Belgium,  \\
FINEP, CNPq, CAPES, FUJB and FAPERJ, Brazil, \\
Czech Ministry of Industry and Trade, GA CR 202/99/1362,\\
Commission of the European Communities (DG XII), \\
Direction des Sciences de la Mati$\grave{\mbox{\rm e}}$re, CEA, France, \\
Bundesministerium f$\ddot{\mbox{\rm u}}$r Bildung, Wissenschaft, Forschung 
und Technologie, Germany,\\
General Secretariat for Research and Technology, Greece, \\
National Science Foundation (NWO) and Foundation for Research on Matter (FOM),
The Netherlands, \\
Norwegian Research Council,  \\
State Committee for Scientific Research, Poland, SPUB-M/CERN/PO3/DZ296/2000,
SPUB-M/CERN/PO3/DZ297/2000 and 2P03B 104 19 and 2P03B 69 23(2002-2004)\\
JNICT--Junta Nacional de Investiga\c{c}\~{a}o Cient\'{\i}fica 
e Tecnol$\acute{\mbox{\rm o}}$gica, Portugal, \\
Vedecka grantova agentura MS SR, Slovakia, Nr. 95/5195/134, \\
Ministry of Science and Technology of the Republic of Slovenia, \\
CICYT, Spain, AEN99-0950 and AEN99-0761,  \\
The Swedish Natural Science Research Council,      \\
Particle Physics and Astronomy Research Council, UK, \\
Department of Energy, USA, DE-FG02-01ER41155, \\
EEC RTN contract HPRN-CT-00292-2002. \\

\newpage
\section*{Appendix A: Results on W polar angle measurements}
\label{appa}

The details of the results for the measurement of the W$^-$ polar angle
differential cross-section are reported in Table~\ref{tab:dsdcost}. For convenience
the integrated luminosities and the lumosity-weighted centre-of-mass
energies in each of the energy bins defined in the text are also provided.

Table~\ref{tab:migration} presents the signal event migration matrix 
determined from the Monte Carlo. The Standard Model W polar angle 
differential distribution is therefore implicitly assumed. 

\begin{table}[htb]
\begin{center}
\begin{tabular}{|c|c|c|c|c|c|c|c|c|c|}
\hline
& & \multicolumn{2}{|c|}{180-184 GeV} & \multicolumn{2}{|c|}{184-194 GeV} & \multicolumn{2}{|c|}{194-204 GeV} & \multicolumn{2}{|c|}{204-210 GeV} \\
\hline
\multicolumn{2}{|c|}{${\cal L}$ (pb$^{-1}$)} & \multicolumn{2}{|c|}{51.63} & \multicolumn{2}{|c|}{178.32} & \multicolumn{2}{|c|}{193.52} & \multicolumn{2}{|c|}{198.59} \\
\hline
\multicolumn{2}{|c|}{weighted $\sqrt{s}$ (GeV)} & \multicolumn{2}{|c|}{182.65} & \multicolumn{2}{|c|}{189.03} & \multicolumn{2}{|c|}{198.46} & \multicolumn{2}{|c|}{205.91} \\
\hline
\multicolumn{2}{|c|}{$\sigma_1$ [-1,-0.8)} & \multicolumn{2}{|c|}{0.715} & \multicolumn{2}{|c|}{0.865} & \multicolumn{2}{|c|}{0.600} & \multicolumn{2}{|c|}{0.275} \\
\hline
$\delta\sigma_1^{stat.}$(meas) & $\delta\sigma_1^{stat.}$(exp) & 0.320 & 0.320 & 0.180 & 0.165 & 0.155 & 0.150 & 0.120 & 0.145 \\
\hline
$\delta\sigma_1^{syst.}$(back) & $\delta\sigma_1^{syst.}$(eff) & 0.045 & 0.020 & 0.040 & 0.020 & 0.025 & 0.015 & 0.020 & 0.015 \\
\hline
\multicolumn{2}{|c|}{$\sigma_2$ [-0.8,-0.6)} & \multicolumn{2}{|c|}{0.795} & \multicolumn{2}{|c|}{0.760} & \multicolumn{2}{|c|}{0.675} & \multicolumn{2}{|c|}{0.590} \\
\hline
$\delta\sigma_2^{stat.}$(meas) & $\delta\sigma_2^{stat.}$(exp) & 0.315 & 0.315 & 0.170 & 0.170 & 0.160 & 0.160 & 0.145 & 0.150 \\
\hline
$\delta\sigma_2^{syst.}$(back) & $\delta\sigma_2^{syst.}$(eff) & 0.025 & 0.025 & 0.020 & 0.020 & 0.015 & 0.020 & 0.015 & 0.020 \\
\hline
\multicolumn{2}{|c|}{$\sigma_3$ [-0.6,-0.4)} & \multicolumn{2}{|c|}{1.175} & \multicolumn{2}{|c|}{0.990} & \multicolumn{2}{|c|}{1.510} & \multicolumn{2}{|c|}{0.575} \\
\hline
$\delta\sigma_3^{stat.}$(meas) & $\delta\sigma_3^{stat.}$(exp) & 0.380 & 0.350 & 0.185 & 0.180 & 0.215 & 0.170 & 0.140 & 0.160 \\
\hline
$\delta\sigma_3^{syst.}$(back) & $\delta\sigma_3^{syst.}$(eff) & 0.020 & 0.035 & 0.020 & 0.035 & 0.015 & 0.030 & 0.010 & 0.025 \\
\hline
\multicolumn{2}{|c|}{$\sigma_4$ [-0.4,-0.2)} & \multicolumn{2}{|c|}{1.365} & \multicolumn{2}{|c|}{0.930} & \multicolumn{2}{|c|}{1.150} & \multicolumn{2}{|c|}{0.930} \\
\hline
$\delta\sigma_4^{stat.}$(meas) & $\delta\sigma_4^{stat.}$(exp) & 0.400 & 0.370 & 0.180 & 0.200 & 0.190 & 0.180 & 0.170 & 0.175 \\
\hline
$\delta\sigma_4^{syst.}$(back) & $\delta\sigma_4^{syst.}$(eff) & 0.015 & 0.035 & 0.015 & 0.035 & 0.015 & 0.035 & 0.010 & 0.035 \\
\hline
\multicolumn{2}{|c|}{$\sigma_5$ [-0.2,0)} & \multicolumn{2}{|c|}{1.350} & \multicolumn{2}{|c|}{1.330} & \multicolumn{2}{|c|}{1.055} & \multicolumn{2}{|c|}{1.000} \\
\hline
$\delta\sigma_5^{stat.}$(meas) & $\delta\sigma_5^{stat.}$(exp) & 0.400 & 0.405 & 0.215 & 0.215 & 0.185 & 0.200 & 0.175 & 0.195 \\
\hline
$\delta\sigma_5^{syst.}$(back) & $\delta\sigma_5^{syst.}$(eff) & 0.015 & 0.040 & 0.015 & 0.040 & 0.015 & 0.035 & 0.015 & 0.035 \\
\hline
\multicolumn{2}{|c|}{$\sigma_6$ [0,0.2)} & \multicolumn{2}{|c|}{1.745} & \multicolumn{2}{|c|}{1.460} & \multicolumn{2}{|c|}{1.635} & \multicolumn{2}{|c|}{1.190} \\
\hline
$\delta\sigma_6^{stat.}$(meas) & $\delta\sigma_6^{stat.}$(exp) & 0.450 & 0.450 & 0.225 & 0.240 & 0.225 & 0.230 & 0.195 & 0.220 \\
\hline
$\delta\sigma_6^{syst.}$(back) & $\delta\sigma_6^{syst.}$(eff) & 0.025 & 0.085 & 0.020 & 0.085 & 0.015 & 0.085 & 0.010 & 0.085 \\
\hline
\multicolumn{2}{|c|}{$\sigma_7$ [0.2,0.4)} & \multicolumn{2}{|c|}{1.995} & \multicolumn{2}{|c|}{1.675} & \multicolumn{2}{|c|}{2.115} & \multicolumn{2}{|c|}{2.120} \\
\hline
$\delta\sigma_7^{stat.}$(meas) & $\delta\sigma_7^{stat.}$(exp) & 0.485 & 0.505 & 0.240 & 0.270 & 0.255 & 0.260 & 0.255 & 0.250 \\
\hline
$\delta\sigma_7^{syst.}$(back) & $\delta\sigma_7^{syst.}$(eff) & 0.015 & 0.050 & 0.015 & 0.050 & 0.010 & 0.045 & 0.010 & 0.045 \\
\hline
\multicolumn{2}{|c|}{$\sigma_8$ [0.4,0.6)} & \multicolumn{2}{|c|}{2.150} & \multicolumn{2}{|c|}{2.630} & \multicolumn{2}{|c|}{3.175} & \multicolumn{2}{|c|}{2.655} \\
\hline
$\delta\sigma_8^{stat.}$(meas) & $\delta\sigma_8^{stat.}$(exp) & 0.510 & 0.580 & 0.300 & 0.320 & 0.320 & 0.310 & 0.290 & 0.300 \\
\hline
$\delta\sigma_8^{syst.}$(back) & $\delta\sigma_8^{syst.}$(eff) & 0.015 & 0.065 & 0.015 & 0.060 & 0.015 & 0.055 & 0.010 & 0.055 \\
\hline
\multicolumn{2}{|c|}{$\sigma_9$ [0.6,0.8)} & \multicolumn{2}{|c|}{4.750} & \multicolumn{2}{|c|}{4.635} & \multicolumn{2}{|c|}{4.470} & \multicolumn{2}{|c|}{4.585} \\
\hline
$\delta\sigma_9^{stat.}$(meas) & $\delta\sigma_9^{stat.}$(exp) & 0.775 & 0.695 & 0.405 & 0.385 & 0.385 & 0.380 & 0.385 & 0.380 \\
\hline
$\delta\sigma_9^{syst.}$(back) & $\delta\sigma_9^{syst.}$(eff) & 0.030 & 0.095 & 0.025 & 0.100 & 0.025 & 0.105 & 0.020 & 0.110 \\
\hline
\multicolumn{2}{|c|}{$\sigma_{10}$ [0.8,1]} & \multicolumn{2}{|c|}{6.040} & \multicolumn{2}{|c|}{5.400} & \multicolumn{2}{|c|}{7.140} & \multicolumn{2}{|c|}{7.290} \\
\hline
$\delta\sigma_{10}^{stat.}$(meas) & $\delta\sigma_{10}^{stat.}$(exp) & 0.895 & 0.850 & 0.455 & 0.490 & 0.500 & 0.505 & 0.505 & 0.520 \\
\hline
$\delta\sigma_{10}^{syst.}$(back) & $\delta\sigma_{10}^{syst.}$(eff) & 0.035 & 0.075 & 0.035 & 0.085 & 0.030 & 0.100 & 0.030 & 0.110 \\
\hline
\end{tabular}
\vspace{0.2cm}
\caption{\label{tab:dsdcost} Differential cross-sections in the 10
angular bins for the four 
energy intervals (see descriptions in Section~\ref{sec:diffxsec}). 
$\sigma_{i}$ indicates the average of 
d[$\sigma_{\W\W}$(BR$_{e\nu}$+BR$_{\mu\nu}$)]/dcos$\theta_{\W^-}$ 
in the $i$-th bin of cos$\theta_{\W^-}$ with width 0.2.
The limits of the bins are also reported in the table.
The values, in each bin, of the 
measured and expected statistical error and of the systematic errors
due to the background and to the efficiencies are reported as well. 
All systematic errors have to be considered correlated in energy and bin.
All values are expressed in pb.}
\end{center}
\end{table}

\begin{table}[htb]
\begin{center}
\begin{tabular}{|c|c c c c c c c c c c|}
\hline
      & gen~1& gen~2&gen~3 &gen~4 &gen~5 &gen~6 &gen~7 &gen~8 &gen~9 &gen~10 \\
\hline
sel~1 & 71.8 & 13.4 &  2.1 &  1.3 &  0.5 &  0.4 &  0.1 &  0.5 &  1.8 &  8.0 \\
sel~2 & 11.4 & 61.7 & 15.4 &  2.7 &  2.1 &  0.8 &  0.6 &  0.7 &  2.4 &  2.2 \\
sel~3 &  1.5 & 12.2 & 57.2 & 16.7 &  4.2 &  1.5 &  1.4 &  1.8 &  2.1 &  1.3 \\
sel~4 &  0.8 &  1.9 & 11.7 & 58.9 & 16.8 &  4.1 &  2.6 &  1.5 &  0.9 &  0.8 \\
sel~5 &  0.2 &  0.5 &  1.7 & 12.8 & 56.4 & 18.9 &  5.0 &  2.4 &  1.1 &  1.0 \\
sel~6 &  0.2 &  0.2 &  0.4 &  1.8 & 12.9 & 58.8 & 18.7 &  3.9 &  1.9 &  1.3 \\
sel~7 &  0.1 &  0.1 &  0.4 &  0.6 &  1.7 & 12.0 & 59.8 & 19.9 &  4.0 &  1.5 \\
sel~8 &  0.0 &  0.1 &  0.1 &  0.3 &  0.5 &  1.7 & 11.6 & 63.3 & 19.4 &  3.2 \\
sel~9 &  0.1 &  0.0 &  0.1 &  0.1 &  0.2 &  0.4 &  0.9 &  9.9 & 70.7 & 17.6 \\
sel~10&  0.1 &  0.0 &  0.0 &  0.0 &  0.1 &  0.1 &  0.2 &  0.7 &  7.4 & 91.4 \\
\hline
\end{tabular}
\vspace{0.2cm}
\caption{\label{tab:migration} Polar angle migration matrix at 200~GeV
for selected events
of the signal, according to the definition described in the text.
The matrix is defined as $M_{ij}=N_{sel i; gen j}/N_{sel i}$, where
$N_{sel i; gen j}$ is the number of events selected in bin $i$ and generated
in bin $j$ and $N_{sel i}$ is the total number of events selected in
bin $i$. The numbers are all expressed in percent.
By construction the rows sum up to 100\%.
The relative errors on the numbers on the diagonal are below 2\%, whereas
outside the diagonal they reach 7\%.}
\end{center}
\end{table}



\newpage

\end{document}